# The MC-QTAIM: Why, How, and When?¶

**17 November 2026**

Shant Shahbazian

Alikhanian National Laboratory (Yerevan Physics Institute), Alikhanian Brothers Street 2, Yerevan 0036, Armenia
E-mail: shant.shahbazyan@aanl.am

## Abstract

The multi-component quantum theory of atoms in molecules (MC-QTAIM) generalizes the original QTAIM framework developed by Richard Bader and colleagues, extending atoms-in-molecules analysis beyond ab initio electronic wavefunctions. MC-QTAIM facilitates the examination of the AIM structure in systems and states that require MC wavefunctions, which describe systems containing multiple distinguishable classes of quantum particles. This advancement broadens the applicability of AIM analysis to a wide range of quantum systems and states that were previously inaccessible. These include exotic molecules, such as positronic species; exotic quantum states, such as superposed proton states in molecules with two- or multi-well potential energy surfaces; and non-adiabatic states of conventional molecules. In this review, I aim to explain the fundamental principles of MC-QTAIM analysis in accessible language and explore some of its potential applications. Additionally, the general formalism for energy partitioning in MC systems, called MC Interacting Quantum Atoms (MC-IQA), is introduced. This scheme unifies the energy partitioning of numerous quantum systems and encompasses the IQA method originally developed by Pendas and colleagues, as well as the previously developed two-component IQA (TC-IQA) scheme—developed by our group to analyze positronic systems—as special cases.

¶ Dedicated to the memory of Dr. Mohammad Mehdi Tehranchi, a man devoted to his fate and his nation.

## 1. Introduction

MC-QTAIM [1,2], stands for the "multi-component quantum theory of atoms in molecules" and represents a generalization of Bader's original QTAIM partitioning methodology [3–7]. Like QTAIM, it is a partitioning method used in bonding analysis that divides molecular properties into atomic contributions using ab initio wavefunctions. What distinguishes MC-QTAIM from QTAIM is that QTAIM analysis can only be applied to electronic wavefunctions, $\Psi_{ele}\left(\vec{r}_{e,1},\vec{r}_{e,2},\ldots,\vec{r}_{e,N_e},\{\sigma_{e,i}\};\{\vec{R}_k\}\right)$, as inputs, which are typically derived from standard ab initio quantum chemical methods applied to molecules [8,9]. In contrast, MC-QTAIM analysis can also utilize "non-electronic" wavefunctions as inputs, as demonstrated in this paper.

Various approximate ab initio electronic wavefunctions are employed in the QTAIM analysis of molecules and condensed phases [10], with Slater determinants being among the simplest examples [8]. Regardless of their complexity or construction method, all these wavefunctions depend on the position vectors, $\vec{r}_{e,i}$, and spin variables, $\sigma_{e,i}$, of all $N_e$ electrons and parametrically on the positions of the clamped nuclei, $\vec{R}_k$. The Born-Oppenheimer (BO) approximation, based on the adiabatic separation of electronic and nuclear dynamics [11–13], is the fundamental principle underlying this framework. Instead of a single time-independent Schrödinger equation (SE) for both electrons and nuclei, two separate SEs arise: the electronic SE, which governs the electrons while treating nuclei as fixed point charges, and the nuclear SE, which governs the motion of nuclei within an effective potential generated by the electrons, known as the potential energy surface (see Chapter 1 in [11]). After separating the center-of-mass translational motion, the nuclear SE is typically expressed in a suitable set of variables describing vibrations, rotations, and their couplings. Although no universally accepted form exists for these variables, various choices have been made, each suited to specific classes of systems [12]; the Watson-Eckart Hamiltonian is perhaps the best known [13]. This review does not delve into the details of the vibrational-rotational Hamiltonian; interested readers may consult comprehensive discussions elsewhere [14–16]. The total adiabatic wavefunction is the product of the electronic and nuclear vibrational-rotational wavefunctions, $\Psi_{ad}=\Psi_{ele}\left(\vec{r}_{e,1},\vec{r}_{e,2},\ldots,\vec{r}_{e,N_e},\{\sigma_{e,i}\};\{\vec{R}_k\}\right)\Psi_{vib-rot}\left(\{\vec{r}_{vib-rot,i}\},\{\sigma_{nuc,j}\}\right)$, each obtained from the solution of the corresponding SE. When the adiabatic separation is no longer valid (vide infra), multiple electronic and vibrational-rotational states may be linearly combined in the Born-Huang (BH) variational expansion [17]. Alternatively, the BO perturbation method, at higher perturbative orders, yields non-adiabatic corrections to the energy and wavefunction [11,18–20]. In such cases, various electronic, vibrational, and rotational states become coupled, and the system cannot be accurately described by a single electronic wavefunction.

Bader began his work at the dawn of the third age of quantum chemistry [21], when the primary challenge was obtaining a quantitatively accurate, chemically precise solution to the electronic SE. The ab initio derivation of vibrational-rotational wavefunctions was not a central focus. This lack of access to ab initio vibrational-rotational wavefunctions resulted in the complete neglect of nuclear motions in QTAIM analysis, limiting the role of nuclei to examining the parametric dependence of the analysis on their positions [22]. However, it is important to emphasize that this issue was not, and is not, unique to QTAIM analysis. To the best of my knowledge, all methods developed for structure and bonding analysis during that era shared this neglect, with only scattered attempts to the contrary, which largely went unnoticed [23–26]. This situation is somewhat understandable, as the chemical community traditionally views bonding analysis as primarily concerned with electrons [27,28]. It is commonly assumed that nuclear motions have no significant effect on structure and bonding. Cases such as “vibrational bonding” [29–32], the enigmatic structure and bonding modes in protonated methane, $CH_5^+$, [33–35], and Bullvalene, $C_{10}H_{10}$, [36,37]—all fluxional molecules—serve as counterexamples. However, these are usually regarded as interesting but rather exceptional curiosities. In his later years, Bader expressed concern about this limitation and believed that nuclear motions should be incorporated into a generalized version of QTAIM analysis (private communications). Setting historical details aside, we are now living in the fourth, or perhaps even fifth, age of quantum chemistry [15,38], in which ab initio solutions of the vibrational-rotational SE are accessible. Therefore, there is no excuse for neglecting the role of nuclear motions in bonding analysis.

Another major advance in recent decades is the development and computational implementation of ab initio “non-Born-Oppenheimer” (nBO) quantum chemical methods [39–41]. Unlike the BO-based methods discussed above, these ab initio approaches do not assume the dynamical decoupling of electrons and nuclei from the outset; instead, they aim to solve the full molecular SE equation. The resulting fully non-adiabatic molecular wavefunctions, formally expressed as $\Psi_{fully-non-ad}\left(\vec{r}_{e,1},\vec{r}_{e,2},...,\vec{r}_{e,N_e},\vec{R}_{nuc,1},\vec{R}_{nuc,2},...,\vec{R}_{nuc,N_{nuc}},\left\{\sigma_{e,i},\sigma_{nuc,j}\right\}\right)$, are mathematically complex and explicitly depend on interparticle distances [40–42]. Moreover, these wavefunctions lack the conceptually appealing separation into electronic, vibrational, and rotational motions. Consequently, they are not amenable to the type of bonding analysis typically performed on electronic wavefunctions [43–48]. However, novel mathematical methods have recently been developed to extract information about the nature of structure and bonding from these wavefunctions, with varying degrees of success [49–52]. In this review, these are termed MC wavefunctions to distinguish them from the usual non-adiabatic wavefunctions derived from BH and BO methods, which assume the adiabatic separation as their starting point. The term "component" in this context refers to the number of "distinguishable" classes of quantum particles in the system. For example, if one solves the full SE for $CH_5^+$ assuming all hydrogen nuclei are protons, the resulting wavefunction is three-component, namely the system comprises electrons, protons, and a carbon nucleus. Note that the isotopic substitution increases the number of components; for instance, if in the above example two protons are replaced by one deuterium and one tritium, $CDTH_3^+$, the system's wavefunction becomes five-component. The inputs of MC-QTAIM analysis are MC wavefunctions, which broadens the range of systems amenable to AIM analysis.

In this review, rather than delving into the technical details of the mathematical foundations of the MC-QTAIM formalism—which have been thoroughly covered in the original papers [53–57] and two recent reviews [1,2]—I aim to provide a general overview of why MC-QTAIM is significant in the fourth and fifth eras of quantum chemistry. The intended audience is the typical quantum chemist seeking to understand the necessity of MC-QTAIM in modern bonding analysis, its applicability, and how to implement it. I will specifically avoid discussing the conceptual challenges associated with AIM analysis in general, as well as certain pitfalls in the consistent chemical interpretation of QTAIM (and MC-QTAIM) analyses. These topics have been extensively addressed elsewhere [58–62]. To enhance accessibility, I have included Q&A subsections at the end of each section to address potential questions about MC-QTAIM. Additionally, some material in the final section has been incorporated to assist program developers in the future general-use implementation of MC-QTAIM analysis by modifying existing quantum chemical packages.

## 2. Why MC-QTAIM?

### 2.1. Motivations

In his concise, unconventional book on applying chemical concepts beyond atoms and molecules, Hefferlin boldly attempts to demonstrate that certain chemical principles are applicable across various areas of physics, including nuclear and particle physics, and even string theory [63]. In a more limited sense, I had a similar idea when I began the project to extend the QTAIM [64]. Why restrict AIM analysis solely to electronic wavefunctions? At least two arguments support a broader application.

First, AIM and chemical bonds appear to be genuine chemical concepts that are independent of how we describe a molecule's quantum states. Consider, for example, a water molecule. Every chemist intuitively describes it as a system composed of three AIM, connected by two bonds. One can use QTAIM analysis to formally derive this intuitive picture from an ab initio computed electronic wavefunction of a water molecule [65]. But how does one extract the same AIM structure and bonding information from an ab initio non-adiabatic wavefunction of a water molecule? How can this question be answered? Another argument arises from considering exotic atoms and molecules. Suppose a fundamental particle such as a positron—a positively charged electron [66]—attaches to a molecule, forming a positronic molecule with a bound ground state. These and many similar exotic systems are not merely theoretical constructs; they are actively produced and studied from a chemical perspective in various laboratories [67–74]. Assume that before positron attachment, the ground state of the molecule can be approximated by an electronic wavefunction, and its AIM structure and bonding modes can be deduced from QTAIM analysis of this wavefunction. After positron attachment, a new molecular bound state is formed [75], albeit with a lifetime on the order of nanoseconds [76]. In this newly formed molecule, the adiabatic separation no longer applies because the electron and positron have the same mass. Thus, the corresponding SE includes the kinetic energy operators and interactions of both particle types simultaneously (vide infra). After solving the equation, the resulting ground-state wavefunction is "two-component" (TC), as it depends on the position vectors and spin variables of both the positron and electrons. What are the AIM structure and the nature of bonds in this new state? How can this question be answered? It seems unreasonable to claim that chemical concepts suddenly cease to apply to the molecule after positron attachment. These two arguments highlight the need to extend QTAIM to extract AIM structures and bonding information from MC wavefunctions. Before addressing this issue, let us consider the nature of MC wavefunctions in more detail.

The conventional molecules, composed of electrons and well-known nuclei, as well as exotic molecules containing additional elementary particles such as positrons or positively and negatively charged muons [77], are all MC systems by their very nature. This implies that, when fine magnetic interactions are neglected, the non-relativistic Hamiltonian for all these systems can be expressed as follows:

$$\hat{H}_{MC} = \sum_{k}^{M}\sum_{i}^{N_k}\left(-\frac{\nabla_{k,i}^2}{2m_k} + v_{ext}\left(\vec{r}_{k,i}\right)\right) + \sum_{k}^{M}\sum_{l>k}^{M}\sum_{i}^{N_k}\sum_{j}^{N_l}\frac{q_k q_l}{\left|\vec{r}_{k,i} - \vec{r}_{l,j}\right|} + \sum_{k}^{M}\sum_{i}^{N_k}\sum_{j>i}^{N_k}\frac{q_k^2}{\left|\vec{r}_{k,i} - \vec{r}_{k,j}\right|} \quad (1)$$

This MC Hamiltonian describes a system composed of several components, $M$ , each containing $N_k$ number of particles characterized by their mass, $m_k$ , and charge, $q_k$ . These particles interact with one another through Coulomb's law and with an arbitrary external field, which must be specified for the system under study. This external field can be an electric field generated by fixed charges, a set of harmonic potentials, or it can be set to zero. It can be readily demonstrated that by adjusting the number, mass, and charge of each component and incorporating the desired external field, any conventional or exotic molecule can be modeled using this Hamiltonian. The aforementioned nBO ab initio methods aim to solve the eigenvalue problem for the MC Hamiltonian, where the resulting ab initio eigenfunctions are fully non-adiabatic MC wavefunctions when the external field is set to zero. For computational tractability, and by applying the adiabatic separation principle to systems with components exhibiting large mass disparities, two simpler Hamiltonians—one for the light components and another for the heavy components—may be derived [11,18,78,79]. As a typical example, consider a system composed of one light component, referred to as component number 1, and one or more heavy components, $M-1$. This scenario corresponds to typical molecules where electrons are much lighter than nuclei, $m_1 \ll m_k,\ k>1$. In this case, all heavy components are replaced by fixed point charges in the Hamiltonian for the light component, $L=\sum_{k=2}^{M} N_k$ ; to emphasize their fixed-point nature, the charges and position vectors of these heavy components are denoted with capital letters hereafter. The resulting single-component (SC) Hamiltonian for component 1 is as follows:

$$\hat{H}_{SC}=\sum_i^{N_1}\left(-\frac{\nabla_{1,i}^2}{2m_1}+v_{ext}\left(\vec{r}_{1,i}\right)\right)+\sum_i^{N_1}\sum_{j>i}^{N_1}\frac{q_1^2}{\left|\vec{r}_{1,i}-\vec{r}_{1,j}\right|}+\sum_i^{N_1}\sum_\alpha^L\frac{Q_\alpha q_1}{\left|\vec{R}_\alpha-\vec{r}_{1,i}\right|}+\sum_\alpha^L\sum_\beta^L\frac{Q_\alpha Q_\beta}{\left|\vec{R}_\alpha-\vec{R}_\beta\right|} \quad (2)$$

This is the electronic Hamiltonian when the mass and charge are fixed to those of the electron and $v_{ext}\left(\vec{r}_{i,k}\right)=0$. By solving the eigenvalue problem, the electronic wavefunctions for any arrangement of nuclei can be obtained. If, instead of one, there are two light components with comparable masses but significantly different from the rest of the components, $m_1 \sim m_2 \ll m_k$, $k>2$,—for example, in positronic molecules—the resulting TC Hamiltonian is as follows:

$$\hat{H}_{TC}=\sum_i^{N_1}\left(-\frac{\nabla_{1,i}^2}{2m_1}+v_{ext}\left(\vec{r}_{1,i}\right)\right)+\sum_i^{N_2}\left(-\frac{\nabla_{2,i}^2}{2m_2}+v_{ext}\left(\vec{r}_{2,i}\right)\right)+\sum_i^{N_1}\sum_{j>i}^{N_1}\frac{q_1^2}{\left|\vec{r}_{1,i}-\vec{r}_{1,j}\right|}+\sum_i^{N_2}\sum_{j>i}^{N_2}\frac{q_2^2}{\left|\vec{r}_{2,i}-\vec{r}_{2,j}\right|}$$

$$+\sum_i^{N_1}\sum_j^{N_2}\frac{q_1 q_2}{\left|\vec{r}_{1,i}-\vec{r}_{2,j}\right|}+\sum_i^{N_1}\sum_\alpha^P\frac{Q_\alpha q_1}{\left|\vec{R}_\alpha-\vec{r}_{1,i}\right|}+\sum_i^{N_2}\sum_\alpha^P\frac{Q_\alpha q_2}{\left|\vec{R}_\alpha-\vec{r}_{2,i}\right|}+\sum_\alpha^P\sum_\beta^P\frac{Q_\alpha Q_\beta}{\left|\vec{R}_\alpha-\vec{R}_\beta\right|} \quad (3)$$

In this Hamiltonian, $P=\sum_{k=3}^{M}N_k$ represents the number of fixed charges. After solving the eigenvalue problem, a TC wavefunction is obtained. By a similar procedure, Hamiltonians can be derived for systems composed of three or more light components. Evidently, various approximate Hamiltonians can be obtained from the parent Hamiltonian, $\hat{H}_{MC}$, by applying the adiabatic separation principle. In each case, the total adiabatic wavefunction is the product of the wavefunction for the lighter particles and that for the heavier components, the latter of which must be derived by solving the corresponding SE. This hierarchical structure of Hamiltonians and wavefunctions is the main focus of this review. The QTAIM analysis is confined, by its construction [6], to the SC wavefunctions, although Bader and coworkers applied QTAIM even more restrictively, only to the electronic wavefunctions. To the best of my knowledge, current commercial packages used for QTAIM analysis also perform the analysis exclusively on ab initio-derived electronic wavefunctions. Nevertheless, as we have demonstrated previously [80,81], QTAIM analysis can also be performed on less familiar SC wavefunctions. Figure 1 schematically depicts the main features of the analysis. The analysis is a “super-algorithm” whose inputs and outputs are the SC wavefunctions and the boundaries and properties of AIM, respectively.

In the first step, the boundaries between AIM must be delineated. This is achieved by deriving the zero-flux surfaces by solving the local zero-flux equation of the one-particle density of the light particle, where $\vec{n}$ is the unit vector normal to the surface [6]:

$$\vec{\nabla}\rho_1\left(\vec{r};\vec{R}_1,\vec{R}_2,...,\vec{R}_L\right).\vec{n}=0,$$

$$\rho_1\left(\vec{r}\right)=N_1\sum_{spins}\int d\vec{r}_2\int d\vec{r}_3...\int d\vec{r}_{N_1}\left|\Psi_{SC}\right|^2$$

$$\left|\Psi_{SC}\right|^2=\Psi^*_{SC}\left(\vec{r}_1,\vec{r}_2,...,\vec{r}_{N_1},\{\sigma_i\};\vec{R}_1,\vec{R}_2,...,\vec{R}_L\right)\Psi_{SC}\left(\vec{r}_1,\vec{r}_2,...,\vec{r}_{N_1},\{\sigma_i\};\vec{R}_1,\vec{R}_2,...,\vec{R}_L\right) \quad (4)$$

For conventional molecules, the light particle is the electron; thus, the relevant density is the electron density, $\Psi_{SC}=\Psi_{ele}$, and $\rho_1=\rho_{ele}$. However, for other species such as pure antimatter molecules, where each particle is replaced by its antiparticle [1], the relevant density is the positron density (interested readers may refer to [81] for additional examples).

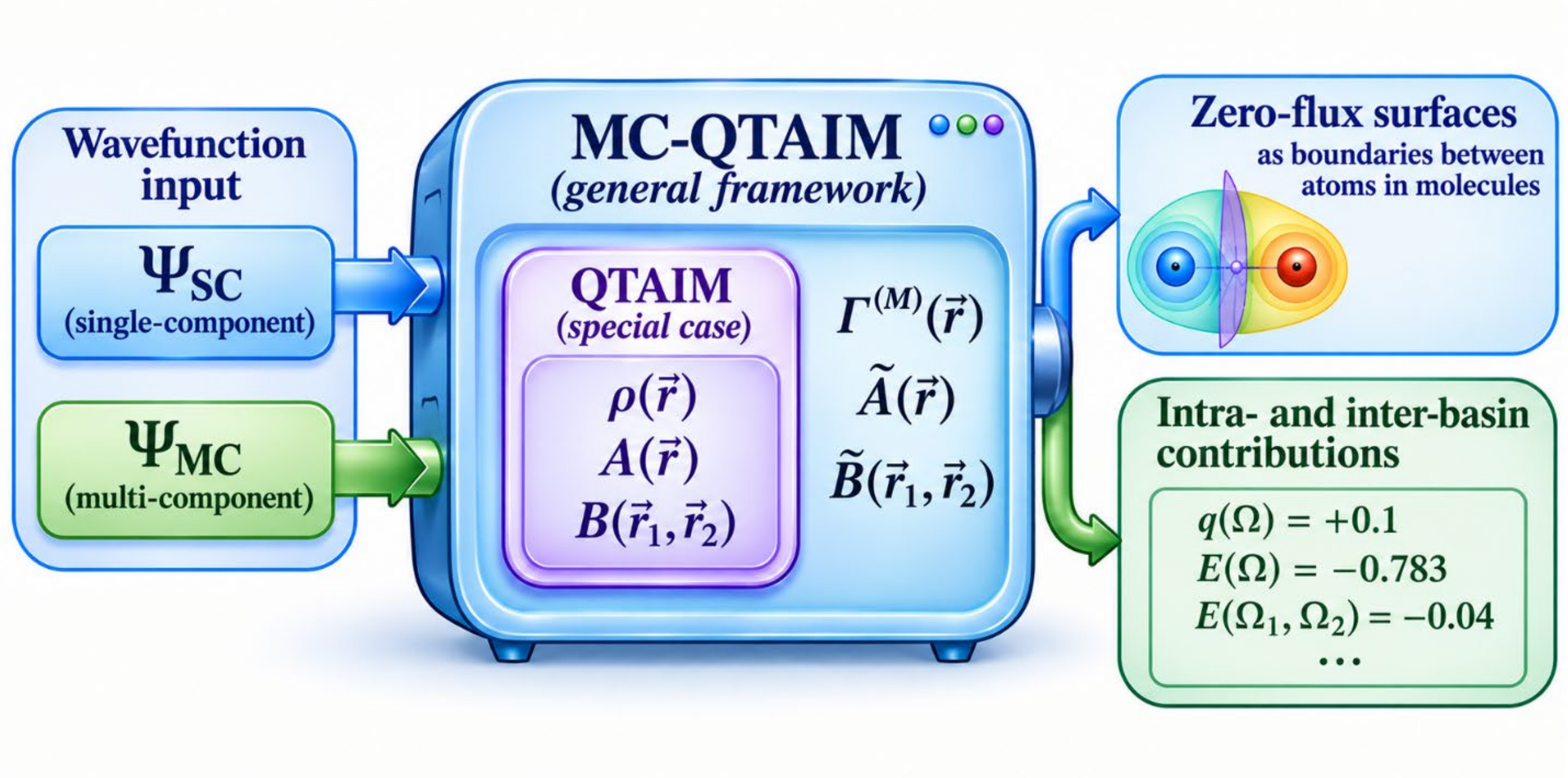

Figure 1- A schematic representation of the relationship between QTAIM and MC-QTAIM. QTAIM is a specific variant of MC-QTAIM that utilizes SC wavefunctions as its input. In contrast, MC-QTAIM employs the more general MC wavefunctions. Both approaches transform these wavefunctions into one-particle densities, one-particle property densities, and two-particle property densities (for details and nomenclature, see the text). The outputs include zero-flux surfaces, which define the boundaries of atomic basins, and properties of AIM. These properties consist of intra- and inter-atomic contributions, such as atomic charges, intra-atomic energy, and inter-basin interaction energy (for details and nomenclature, see the text).

The boundaries between AIM are defined by zero-flux surfaces that do not intersect the nuclei [6,82]. The volumes enclosed by these boundaries, each usually containing a single nucleus, are called atomic basins. An atomic basin, denoted hereafter as $\Omega$, may have either a finite or infinite volume depending on whether it is completely enclosed by zero-flux surfaces—typically the case for certain atoms in cage-like molecules and crystals [10]—or only partially enclosed, which is usually observed in molecules lacking outer boundaries [6].

In the second step, the contribution of each atomic basin to molecular properties is calculated by introducing one- and two-particle property densities [6]. Bader provided the methods for constructing these densities [6], which I have summarized in two recent reviews [1,2]; therefore, I do not revisit the details here. Briefly, for one-particle properties—such as kinetic energy or electric dipole moment—represented by Hermitian operators, $\hat{A}=\sum_{i}^{N_1}\hat{a}_i$, the property density is constructed as follows:

$$A_1(\vec{r})=\hat{a}_1\rho_1^{(1)}(\vec{r}_1,\vec{r}_1')\Big|_{\vec{r}_1=\vec{r}_1'},$$

$$\rho_1^{(1)}(\vec{r}_1,\vec{r}_1')=N_1\sum_{spins}\int d\vec{r}_2\int d\vec{r}_3\ldots\int d\vec{r}_{N_1}\Gamma_1^{(1)}$$

$$\Gamma_1^{(1)}=\Psi_{SC}^{*}\left(\vec{r}_1',\vec{r}_2,\ldots,\vec{r}_{N_1},\{\sigma_i\};\vec{R}_1,\vec{R}_2,\ldots,\vec{R}_L\right)\Psi_{SC}\left(\vec{r}_1,\vec{r}_2,\ldots,\vec{r}_{N_1},\{\sigma_i\};\vec{R}_1,\vec{R}_2,\ldots,\vec{R}_L\right) \quad (5)$$

In these expressions, $\rho_1^{(1)}(\vec{r},\vec{r}')$ is the reduced first-order density matrix [83], and the notation $\rho_1^{(1)}(\vec{r},\vec{r}')\Big|_{\vec{r}=\vec{r}'}$ indicates that after the operator acts, $\hat{a}$, the prime on the primed vector is removed. Integrating the corresponding one-particle property density over the volume of an atomic basin yields the contribution of that basin to the property, $A_1(\Omega_p)=\int_{\Omega_p} d\vec{r}A_1(\vec{r})$. For a system with $S$ atomic basins, the sum of all atomic contributions gives the expectation value of that property for the entire molecule $\langle\hat{A}_1\rangle=\int d\tau\ \Psi_{SC}^{*}\hat{A}_1\Psi_{SC}=\sum_{p}^{S}A_1(\Omega_p)$ (hereafter, $d\tau$ implies integration over all variables of all particles involved in the wavefunction). For two-particle properties, such as the electron-electron repulsion term, $\hat{B}=\sum_{i}^{N_1}\sum_{j>i}^{N_1}\hat{b}_{ij}$, the property density is given by:

$$B_{11}(\vec{r}_1,\vec{r}_2)=\hat{b}_{12}\rho_{11}^{(2)}(\vec{r}_1,\vec{r}_2),$$

$$\rho_{11}^{(2)}\left(\vec{r}_1,\vec{r}_2\right)=\left(1/2\right)N_1\left(N_1-1\right)\sum_{spins}\int d\vec{r}_3\int d\vec{r}_4...\int d\vec{r}_{N_1}\left|\Psi_{SC}\right|^2 \quad (6)$$

In these expressions, $\rho_{11}^{(2)}\left(\vec{r}_1,\vec{r}_2\right)$ is the diagonal part of the reduced second-order density matrix [83]. Since two-particle operators usually do not include derivatives, the prime notation used in Eq. (5) is not necessary here. From this two-particle density, the intra- and inter-basin properties can be derived as follows: $B_{11}\left(\Omega_p\right)=\int_{\Omega_p} d\vec{r}_1\int_{\Omega_p} d\vec{r}_2 B_{11}\left(\vec{r}_1,\vec{r}_2\right)$, $B_{11}\left(\Omega_p,\Omega_q\right)=\int_{\Omega_p} d\vec{r}_1\int_{\Omega_q} d\vec{r}_2 B_{11}\left(\vec{r}_1,\vec{r}_2\right)$, respectively. The sum over all intra- and inter-atomic basins yields the expectation value of the property for the molecule, $\left\langle\hat{B}_{11}\right\rangle=\int d\tau\ \Psi_{SC}^{*}\hat{B}_{11}\Psi_{SC}=\sum_{p}^{S}B_{11}\left(\Omega_p\right)+\sum_{p}^{S}\sum_{q\neq p}^{S}B_{11}\left(\Omega_p,\Omega_q\right)$. Note that in this equation, due to the indistinguishability of the particles, $B_{11}\left(\Omega_p,\Omega_q\right)=B_{11}\left(\Omega_q,\Omega_p\right)$, thus, $\sum_{p}^{S}\sum_{q\neq p}^{S}B_{11}\left(\Omega_p,\Omega_q\right)=2\sum_{p}^{S}\sum_{q>p}^{S}B_{11}\left(\Omega_p,\Omega_q\right)$. Basin properties are used to interpret various molecular phenomena [10,65].

To perform AIM analysis on two-, three-, and, more generally, MC wavefunctions, these two steps must be extended to account for the contributions of all the light components. At first, it may seem that such an extension can be implemented in various ways. However, as discussed in the subsequent sections, this is not the case; only a single formalism emerges after carefully generalizing the principles Bader used to develop QTAIM.

### 2.2. Q&A

1) *Does the QTAIM formalism explicitly depend on the specific form of the employed Hamiltonian?*

Yes, as is evident from the discussions in the previous subsection, the QTAIM formalism has been derived for the SC Hamiltonian, with the additional constraint of $v_{ext}\left(\vec{r}_{i,k}\right)=0$. Bader claimed that this Hamiltonian is "universally" sufficient for describing all chemically interesting systems and phenomena [84]. Nevertheless, even excluding exotic systems, many chemically significant phenomena in molecules are intrinsically non-adiabatic, such as the Jahn-Teller effect, various vibronic couplings, and proton-coupled electron transfer [85–87]. In these cases, the electronic and vibrational-rotational motions are coupled, and the system or phenomenon cannot be accurately described by any SC Hamiltonian. It is important to emphasize that this issue is not unique to the QTAIM formalism; most, if not all, available methods for bonding analysis can only be applied to SC wavefunctions. Furthermore, this dependence on the Hamiltonian is not limited to QTAIM—other methods developed and used in quantum chemistry exhibit similar characteristics. The fundamental Hohenberg-Kohn theorems of density functional theory provide prime examples, as their applicability changes significantly for stationary states, time-dependent states, relativistic Hamiltonians, and MC systems (see [88], particularly chapter 9, for a succinct but informative discussion).

To encompass a broad range of chemically significant systems and phenomena, QTAIM must be extended to include stationary MC states and time-dependent states, which Bader mentioned but did not elaborate on in detail [6,89], as well as relativistic states [90–93]. Only a unified AIM analysis method capable of addressing all these states—and potentially others—can claim universal applicability. Currently, we are far from achieving such a unified formalism.

2) *Is it possible to use the vibrational-rotational wavefunction to calculate the contribution of atomic basins to nuclear-motion-related properties?*

In principle, yes, although this has never been fully developed in detail. As mentioned previously, QTAIM analysis utilizes only the one-electron density and one- and two-electron property densities derived from the electronic wavefunction. Nevertheless, if a nuclear motion-related property density can be constructed using the vibrational-rotational wavefunction, then, by integration, the contribution of each atomic basin to that property can be determined. For the one-particle operators related to vibrational motion, we have proposed the following equations: $A_{vib}\left(\Omega_p\right)=\int_{\Omega_p} d\vec{r} A_{vib}\left(\vec{r}\right)$ and $\left\langle \hat{A}_{vib}\right\rangle=\int d\tau\ \Psi^*_{vib}\hat{A}_{vib}\Psi_{vib}=\sum_{p}^{S} A_{vib}\left(\Omega_p\right)$ [94], and the total contribution of each atomic basin to a property is the sum of its electronic and vibrational components: $A_{total}\left(\Omega_p\right)=A_{ele}\left(\Omega_p\right)+A_{vib}\left(\Omega_p\right)$. The key step toward a workable formalism is a proper method for constructing the property densities. We have tentatively proposed such a method for the energy density of a set of uncoupled vibrating "distinguishable" nuclei using the MC-QTAIM formalism [94]. This approach is justified since the corresponding Hamiltonian is a special case of $\hat{H}_{MC}$. After basin integration, the contribution of each atom to the zero-point vibrational energy is consistently derived [94]. However, a general method for constructing property densities for a set of vibrationally coupled indistinguishable nuclei, whether fermions or bosons, has yet to be developed. A properly developed formalism would naturally include “isotope effects” [95], which are absent from standard QTAIM analysis, since nuclear masses enter the vibrational Hamiltonian, affecting its eigenfunctions, $\Psi_{vib}$, and making $A_{vib}\left(\Omega_p\right)$ distinct for various isotopes. Finally, the role of rotational motions must also be incorporated into this formalism. Beyond rigid rotations [96], the centrifugal forces operative in non-rigid rotations deform the molecular geometry and alter the mean relative distances between vibrating nuclei [97]. This effect is particularly pronounced in highly excited vibrational-rotational states, where the structure and bonding change drastically compared to the ground vibrational-rotational state [98]. Therefore, to fully incorporate all these effects, an extended formalism must be developed whose inputs are both $\Psi_{ele}$ and $\Psi_{vib-rot}$. Currently, we are far from achieving such an extended QTAIM analysis.

Nevertheless, any formalism based on the adiabatic approximation has the drawback that the atomic basin boundaries are independent of the vibrational-rotational states. Basin integration is performed using zero-flux surfaces derived from the zero-flux equation of the electron density, and nuclear densities do not influence the shape of the atomic basin. As demonstrated in the following sections, the MC-QTAIM formalism incorporates nuclear one-particle densities directly in defining the boundaries of AIM, thereby overcoming this limitation.

## 3. How MC-QTAIM?

### 3.1. Colorful electrons

The development of MC-QTAIM was a complex and non-linear process; interested readers may refer to the original literature for detailed accounts [53–57,99–102]. To avoid retracing this historical path, I will attempt to justify the MC-QTAIM formalism using the metaphor of "colorful electrons." Although this approach is less rigorous and differs from the original analytical methods, it is more accessible. The concept of colorful electrons, while not explicitly framed in current terminology, has previously been employed to assign a wavefunction to the Promolecule model [103].

Let us assume that, in addition to mass and charge, electrons possess a new (fictitious) property called color, which can take one of two values: red or blue. Measuring this property allows us to distinguish between electrons. Assume that color does not induce new interactions between electrons and does not change due to electromagnetic interactions. Furthermore, the electromagnetic laws are identical for both red and blue electrons. The TC wavefunction of a molecule composed of $N_r$ red and $N_b$ blue electrons is: $\Phi_{TC}\left(\vec{r}_{r,1},\vec{r}_{r,2},...,\vec{r}_{r,N_r},\vec{r}_{b,1},\vec{r}_{b,2},...,\vec{r}_{b,N_b},\left\{\sigma_{r,i},\sigma_{b,j}\right\}\right)$. This wavefunction exhibits fermionic exchange symmetry and changes sign only when the position and spin variables of two electrons within the same group (red or blue) are exchanged. This implies that Pauli's exclusion principle applies separately to electrons within each group but not between groups [103]. The simplest example of such a wavefunction is a product of two Slater determinants, one for red electrons and one for blue electrons. Interestingly, since the charge and mass of these two groups of electrons are identical and color does not affect the interaction terms, the TC Hamiltonian of this system, Eq. (3), effectively reduces to the SC Hamiltonian, Eq. (2), by equating $N_1 = N_r + N_b$. Nevertheless, as mentioned, the eigenfunctions of this Hamiltonian differ from those of the conventional SC electronic system due to differences in exchange symmetry. This represents the closest approximation to an SC system while maintaining a genuine TC wavefunction. Evidently, the QTAIM analysis cannot be applied directly to this TC system. How, then, should QTAIM be extended to apply to this system?

To begin, we need to delineate the boundaries between AIM. At this stage, QTAIM utilizes the one-electron density; however, the TC system contains two distinguishable types of electrons. Therefore, there are two distinct one-electron densities:

$$\rho_r\left(\vec{r}\right) = N_r \sum_{spins} \int d\vec{r}_{r,2} \int d\vec{r}_{r,3} ... \int d\vec{r}_{r,N_r} \int d\vec{r}_{b,1} \int d\vec{r}_{b,2} ... \int d\vec{r}_{b,N_b} \left|\Phi_{TC}\right|^2$$

$$\rho_b\left(\vec{r}\right) = N_b \sum_{spins} \int d\vec{r}_{r,1} \int d\vec{r}_{r,2} ... \int d\vec{r}_{r,N_r} \int d\vec{r}_{b,2} \int d\vec{r}_{b,2} ... \int d\vec{r}_{b,N_b} \left|\Phi_{TC}\right|^2 \qquad (7)$$

We have no reason to exclude red or blue electrons from the analysis, and the rational expectation is that a combination of these two densities, $\tilde{\rho}\left(\vec{r}\right)$, should be used in an extended zero-flux equation to derive the atomic boundaries. To gain insight into the proper form of this combination, we may consider the asymptotic cases. When one of the colors is absent from the system, the following results are easily deducible: $\left(N_r = N_1, N_b = 0\right) \Rightarrow \tilde{\rho}\left(\vec{r}\right) = \rho_r\left(\vec{r}\right)$, $\left(N_b = N_1, N_r = 0\right) \Rightarrow \tilde{\rho}\left(\vec{r}\right) = \rho_b\left(\vec{r}\right)$. In the other asymptotic case, if the color property suddenly disappears from the system and the electrons become indistinguishable, we expect: $\tilde{\rho}\left(\vec{r}\right) = \rho_{ele}\left(\vec{r}\right)$. The linear combination is the simplest form consistent with these asymptotic conditions, $\tilde{\rho}(\vec{r}) = \rho_r(\vec{r}) + \rho_b(\vec{r})$. This line of reasoning can be easily extended to a multi-color system of electrons, $\tilde{\rho}(\vec{r}) = \sum_k^M \rho_k(\vec{r})$, and the tentative zero-flux equation is: $\vec{\nabla}\tilde{\rho}(\vec{r}) \cdot \vec{n} = 0$. The true test of this equation is whether it can yield reasonable atomic boundaries when applied to systems with colorful electrons. When applied to a large set of Promolecule hydrocarbons, we have shown that the equation works surprisingly well [104,105]. This success demonstrates that the combination is a reliable guiding principle for QTAIM's effective generalization to MC systems. We may also apply the combination principle to the second step of QTAIM analysis.

For the one-particle properties, $\hat{A} = \sum_i^{N_r} \hat{a}_{r,i} + \sum_j^{N_b} \hat{a}_{b,j}$, the same rationale dictates the following combination, $\tilde{A}(\vec{r}) = A_r(\vec{r}) + A_b(\vec{r})$, where:

$$A_r(\vec{r}_r) = \hat{a}_r \rho_r^{(1)}(\vec{r}_r, \vec{r}_r')\Big|_{\vec{r}_r = \vec{r}_r'},$$

$$\rho_r^{(1)}(\vec{r}_r, \vec{r}_r') = N_r \sum_{spins} \int d\vec{r}_{r,2} \int d\vec{r}_{r,3} \ldots \int d\vec{r}_{r,N_r} \int d\vec{r}_{b,1} \int d\vec{r}_{b,2} \ldots \int d\vec{r}_{b,N_b} \Gamma_r^{(1)}$$

$$\Gamma_r^{(1)} = \Phi_{TC}^*\left(\vec{r}_{r,1}', \ldots, \vec{r}_{b,N_b}, \{\sigma_{r,i}, \sigma_{b,j}\}\right) \Phi_{TC}\left(\vec{r}_{r,1}, \ldots, \vec{r}_{b,N_b}, \{\sigma_{r,i}, \sigma_{b,j}\}\right)$$

$$A_b(\vec{r}_b) = \hat{a}_b \rho_b^{(1)}(\vec{r}_b, \vec{r}_b')\Big|_{\vec{r}_b = \vec{r}_b'},$$

$$\rho_b^{(1)}(\vec{r}_b, \vec{r}_b') = N_b \sum_{spins} \int d\vec{r}_{r,1} \int d\vec{r}_{r,2} \ldots \int d\vec{r}_{r,N_r} \int d\vec{r}_{b,2} \int d\vec{r}_{b,3} \ldots \int d\vec{r}_{b,N_b} \Gamma_b^{(1)}$$

$$\Gamma_b^{(1)} = \Phi_{TC}^*\left(\vec{r}_{b,1}', \ldots, \vec{r}_{r,N_r}, \{\sigma_{r,i}, \sigma_{b,j}\}\right) \Phi_{TC}\left(\vec{r}_{b,1}, \ldots, \vec{r}_{r,N_r}, \{\sigma_{r,i}, \sigma_{b,j}\}\right) \qquad (8)$$

The basin properties are derived by integrating the combined density, $\tilde{A}(\Omega_p) = \int_{\Omega_p} d\vec{r}\, \tilde{A}(\vec{r}) = A_r(\Omega_p) + A_b(\Omega_p)$, and the sum of the basin properties yields the expectation value for the entire system, $\langle \hat{A} \rangle = \int d\tau\ \Phi_{TC}^* \hat{A} \Phi_{SC} = \sum_p^S \tilde{A}(\Omega_p)$. For two-particle properties, $\hat{B} = \sum_i^{N_r} \sum_{j>i}^{N_r} \hat{b}_{r,ij} + \sum_k^{N_b} \sum_{l>k}^{N_b} \hat{b}_{b,kl} + \sum_i^{N_r} \sum_k^{N_b} \hat{b}_{rb,ik}$, there are intra-group and inter-group operators, and the property densities are defined for each type of operator:

$$B_{rr}(\vec{r}_{r,1}, \vec{r}_{r,2}) = \hat{b}_{r,12} \rho_{rr}^{(2)}(\vec{r}_{r,1}, \vec{r}_{r,2}),$$

$$\rho_{rr}^{(2)}\left(\vec{r}_{r,1},\vec{r}_{r,2}\right)=\left(1/2\right)N_r\left(N_r-1\right)\sum_{spins}\int d\vec{r}_{r,3}...\int d\vec{r}_{r,N_r}\int d\vec{r}_{b,1}...\int d\vec{r}_{b,N_b}\left|\Phi_{TC}\right|^2$$

$$B_{bb}\left(\vec{r}_{b,1},\vec{r}_{b,2}\right)=\hat{b}_{b,12}\rho_{bb}^{(2)}\left(\vec{r}_{b,1},\vec{r}_{b,2}\right),$$

$$\rho_{bb}^{(2)}\left(\vec{r}_{b,1},\vec{r}_{b,2}\right)=\left(1/2\right)N_b\left(N_b-1\right)\sum_{spins}\int d\vec{r}_{r,1}...\int d\vec{r}_{r,N_r}\int d\vec{r}_{b,3}...\int d\vec{r}_{b,N_b}\left|\Phi_{TC}\right|^2$$

$$B_{rb}\left(\vec{r}_{r,1},\vec{r}_{b,1}\right)=\hat{b}_{rb,11}\rho_{rb}^{(2)}\left(\vec{r}_{r,1},\vec{r}_{b,1}\right),$$

$$\rho_{rb}^{(2)}\left(\vec{r}_{r,1},\vec{r}_{b,1}\right)=N_rN_b\sum_{spins}\int d\vec{r}_{r,2}...\int d\vec{r}_{r,N_r}\int d\vec{r}_{b,2}...\int d\vec{r}_{b,N_b}\left|\Phi_{TC}\right|^2 \qquad (9)$$

A new entity, the pair density, $\rho_{rb}^{(2)}\left(\vec{r}_{r,1},\vec{r}_{b,1}\right)$, emerges from considering inter-group operators and has no counterpart in the SC systems. In particular, whereas $\rho_{rr}^{(2)}\left(\vec{r}_{r,1},\vec{r}_{r,2}\right)$ and $\rho_{bb}^{(2)}\left(\vec{r}_{b,1},\vec{r}_{b,2}\right)$ are symmetric with respect to the exchange of the position vectors of the two particles, this symmetry does not hold for the pair density. The rationale for combining densities also applies to two-particle properties, $\tilde{B}_{rb}\left(\vec{r}_1,\vec{r}_2\right)=B_{rr}\left(\vec{r}_1,\vec{r}_2\right)+B_{bb}\left(\vec{r}_1,\vec{r}_2\right)+B_{rb}\left(\vec{r}_1,\vec{r}_2\right)$. Double integrations over the atomic basins yield the intra- and inter-basin contributions for each group of electrons: $\tilde{B}_{rb}\left(\Omega_p,\Omega_p\right)=\int_{\Omega_p}d\vec{r}_1\int_{\Omega_p}d\vec{r}_2\tilde{B}_{rb}\left(\vec{r}_1,\vec{r}_2\right)=B_{rr}\left(\Omega_p,\Omega_p\right)+B_{bb}\left(\Omega_p,\Omega_p\right)+B_{rb}\left(\Omega_p,\Omega_p\right)$, $\tilde{B}_{rb}\left(\Omega_p,\Omega_q\right)=\int_{\Omega_p}d\vec{r}_1\int_{\Omega_q}d\vec{r}_2\tilde{B}_{rb}\left(\vec{r}_1,\vec{r}_2\right)=B_{rr}\left(\Omega_p,\Omega_q\right)+B_{bb}\left(\Omega_p,\Omega_q\right)+B_{rb}\left(\Omega_p,\Omega_q\right)$, and the sum over basin properties yields the expectation value for the total system: $\left\langle\hat{B}\right\rangle=\int d\tau\ \Phi_{TC}^{*}\hat{B}\Phi_{SC}=\sum_{p}^{S}\tilde{B}_{rb}\left(\Omega_p,\Omega_p\right)+\sum_{p}^{S}\sum_{q\neq p}^{S}\tilde{B}_{rb}\left(\Omega_p,\Omega_q\right)$. Note that in this equation, due to the intra-group indistinguishability of red and blue electrons, certain equalities hold, such as $B_{rr}\left(\Omega_p,\Omega_q\right)=B_{rr}\left(\Omega_q,\Omega_p\right)$, and thus $\sum_{p}^{S}\sum_{q\neq p}^{S}B_{rr}\left(\Omega_p,\Omega_q\right)=2\sum_{p}^{S}\sum_{q>p}^{S}B_{rr}\left(\Omega_p,\Omega_q\right)$, $\sum_{p}^{S}\sum_{q\neq p}^{S}B_{bb}\left(\Omega_p,\Omega_q\right)=2\sum_{p}^{S}\sum_{q>p}^{S}B_{bb}\left(\Omega_p,\Omega_q\right)$. However, the inter-group distinguishability of the red and blue electrons implies that, in general, $B_{rb}\left(\Omega_p,\Omega_q\right)\neq B_{rb}\left(\Omega_q,\Omega_p\right)$. Extension of these equations to a multi-color system of electrons is straightforward through a linear combination of all the densities.

$$\left\langle \hat{A} \right\rangle = \int d\tau \; \Phi_{MC}^{*} \hat{A} \Phi_{MC} = \sum_{p}^{S} \tilde{A}\left(\Omega_p\right), \quad \tilde{A}\left(\Omega_p\right) = \int_{\Omega_p} d\vec{r} \tilde{A}\left(\vec{r}\right),$$

$$\tilde{A}\left(\vec{r}\right) = \sum_{k}^{M} A_k\left(\vec{r}\right), \qquad A_k\left(\vec{r}_k\right) = \hat{a}_k \rho_k^{(1)}\left(\vec{r}_k, \vec{r}_k'\right)\Big|_{\vec{r}_k = \vec{r}_k'},$$

$$\rho_k^{(1)}\left(\vec{r}_k, \vec{r}_k'\right) = N_k \sum_{spins} \int d\vec{r}_{1,1} \int d\vec{r}_{1,2} \ldots \int d\vec{r}_{1,N_1} \ldots \int d\vec{r}_{k,2} \int d\vec{r}_{k,3} \ldots \int d\vec{r}_{k,N_k} \ldots$$

$$\ldots \int d\vec{r}_{M,1} \int d\vec{r}_{M,2} \ldots \int d\vec{r}_{M,N_M} \Gamma_k^{(1)} \qquad (10)$$

And:

$$\left\langle \hat{B} \right\rangle = \int d\tau \; \Phi_{MC}^{*} \hat{B} \Phi_{MC} = \sum_{p}^{S} \tilde{B}\left(\Omega_p, \Omega_p\right) + \sum_{p}^{S} \sum_{q \neq p}^{S} \tilde{B}\left(\Omega_p, \Omega_q\right),$$

$$\tilde{B}\left(\Omega_p, \Omega_p\right) = \int_{\Omega_p} d\vec{r}_1 \int_{\Omega_p} d\vec{r}_2 \tilde{B}\left(\vec{r}_1, \vec{r}_2\right), \quad \tilde{B}\left(\Omega_p, \Omega_q\right) = \int_{\Omega_p} d\vec{r}_1 \int_{\Omega_q} d\vec{r}_2 \tilde{B}\left(\vec{r}_1, \vec{r}_2\right)$$

$$\tilde{B}\left(\vec{r}_1, \vec{r}_2\right) = \sum_{k}^{M} \sum_{l \geq k}^{M} B_{kl}\left(\vec{r}_1, \vec{r}_2\right), \quad B_{kk}\left(\vec{r}_{k,1}, \vec{r}_{k,2}\right) = \hat{b}_{k,12} \rho_{kk}^{(2)}\left(\vec{r}_{k,1}, \vec{r}_{k,2}\right),$$

$$B_{kl}\left(\vec{r}_{k,1}, \vec{r}_{l,1}\right) = \hat{b}_{kl,11} \rho_{kl}^{(2)}\left(\vec{r}_{k,1}, \vec{r}_{l,1}\right),$$

$$\rho_{kk}^{(2)}\left(\vec{r}_{k,1}, \vec{r}_{k,2}\right) = \left(1/2\right) N_k \left(N_k - 1\right) \sum_{spins} \int d\vec{r}_{1,1} \int d\vec{r}_{1,2} \ldots \int d\vec{r}_{1,N_1} \ldots \int d\vec{r}_{k,3} \int d\vec{r}_{k,4} \ldots \int d\vec{r}_{k,N_k} \ldots$$

$$\ldots \int d\vec{r}_{M,1} \int d\vec{r}_{M,2} \ldots \int d\vec{r}_{M,N_M} \left|\Phi_{MC}\right|^2,$$

$$\rho_{kl}^{(2)}\left(\vec{r}_{k,1}, \vec{r}_{l,1}\right) = N_k N_l \sum_{spins} \int d\vec{r}_{1,1} \int d\vec{r}_{1,2} \ldots \int d\vec{r}_{1,N_1} \ldots \int d\vec{r}_{k,2} \int d\vec{r}_{k,3} \ldots \int d\vec{r}_{k,N_k} \ldots \int d\vec{r}_{l,2} \ldots \int d\vec{r}_{l,N_l} \ldots$$

$$\ldots \int d\vec{r}_{M,1} \int d\vec{r}_{M,2} \ldots \int d\vec{r}_{M,N_M} \left|\Phi_{MC}\right|^2 \qquad (11)$$

The formalism developed thus far is complete. The most significant remaining question is what happens to this formalism when, instead of fictitious colorful electrons, the target is a molecular system composed of a set of real, distinguishable components.

**3.2. Beyond colorful electrons: The general formalism**

In the developed formalism, each colorful electron contributes equally to the combined densities. However, if the components have different properties, such as masses and charges, it becomes possible to "scale" their contributions. This scaling can be achieved by generalizing the principle of density combination. For a TC system, where component 1 has a smaller mass than component 2, the density combination is generalized as follows: $\tilde{\rho}\left(\vec{r}\right) = \alpha_1 \rho_1\left(\vec{r}\right) + \alpha_2 \rho_2\left(\vec{r}\right)$. There is no straightforward intuitive argument to deduce the scaling factors, $\alpha_1$ and $\alpha_2$, but they appear to depend on the intrinsic properties of the components, $\alpha = \alpha(m, q, ..)$. Only through mathematically complex procedures—by generalizing the subsystem variational method [56,101,106] and the regional/local virial theorem to MC systems [54,56], it been shown that: $\alpha_k = m_1/m_k$. The generalization to MC systems is as follows [56]:

$$\Gamma^{(M)}(\vec{r}) = \rho_1(\vec{r}) + \sum_{k=2}^{M} (m_1/m_k)\rho_k(\vec{r}) \tag{12}$$

$\Gamma^{(M)}(\vec{r})$ is referred to as the Gamma density and $\rho_1(\vec{r})$ represents the one-particle density of the lightest component of the system. The mass scaling is intuitively appealing because a more massive component contributes less to the Gamma density and has a reduced effect on the interatomic boundary. A detailed analytical study of the asymptotic behavior of Gamma demonstrates that if all masses except that of component 1 tend to infinity, all the one-particle densities in the summation above, except for that of component 1, $\rho_1(\vec{r})$, become Dirac delta functions [54]. However, the mass scaling ensures that, in this asymptotic limit, Gamma reduces to the density of component 1, $\rho_1(\vec{r})$, and the system effectively becomes SC. It can be shown analytically that the same behavior occurs for the asymptotic behavior of property densities [54]. This implies that MC-QTAIM encompasses QTAIM as an asymptotic case, as illustrated schematically in Figure 1. For the property densities, it has been shown that the above scaling is unnecessary [54]; intuitively, this is justifiable since masses and charges are inherently incorporated into the definitions of property densities. Accordingly, Eqs. (10) and (11) are used directly for partitioning the properties of a general MC system, with the only modification being the replacement of $\Phi_{MC}$ wavefunction by $\Psi_{MC}$, which is an eigenfunction of the general MC Hamiltonian introduced in Eq. (1). The zero-flux surfaces, which serve as boundaries between AIM, are determined by solving the local zero-flux equation of the Gamma density: $\vec{\nabla}\Gamma^{(M)}(\vec{r}).\vec{n} = 0$. The Gamma density and its gradient vector field, $\vec{\nabla}\Gamma^{(M)}(\vec{r})$, typically exhibit a simple topography for conventional molecules. Figure 2 schematically depicts, as an example, the Gamma density and its gradient for a typical three-atom species. In this system, the two heavy nuclei are treated as clamped point charges in the Hamiltonian, while the light nucleus and electrons are treated as quantum particles. The Gamma density is composed of the electron and nuclear densities, $\Gamma^{(2)}(\vec{r}) = \rho_1(\vec{r}) + (m_e/m_{nuc})\rho_2(\vec{r})$, and since the nuclear density, $\rho_2(\vec{r})$, is mainly localized in a small region of space, the electron density, $\rho_1(\vec{r})$, dominates the topography in most regions. Even in regions with localized nuclear density, the mass scaling, $m_e/m_{nuc} < 10^{-3}$, largely attenuates its influence on shaping the Gamma density. The Gamma density exhibits three maxima, and its vector field consists of three basins of attraction, each basin containing gradient paths that converge to a global attractor, a (3, -3) critical point (CP). The classification scheme of CPs follows that of QTAIM and is referred to as topological analysis [6,60,107].

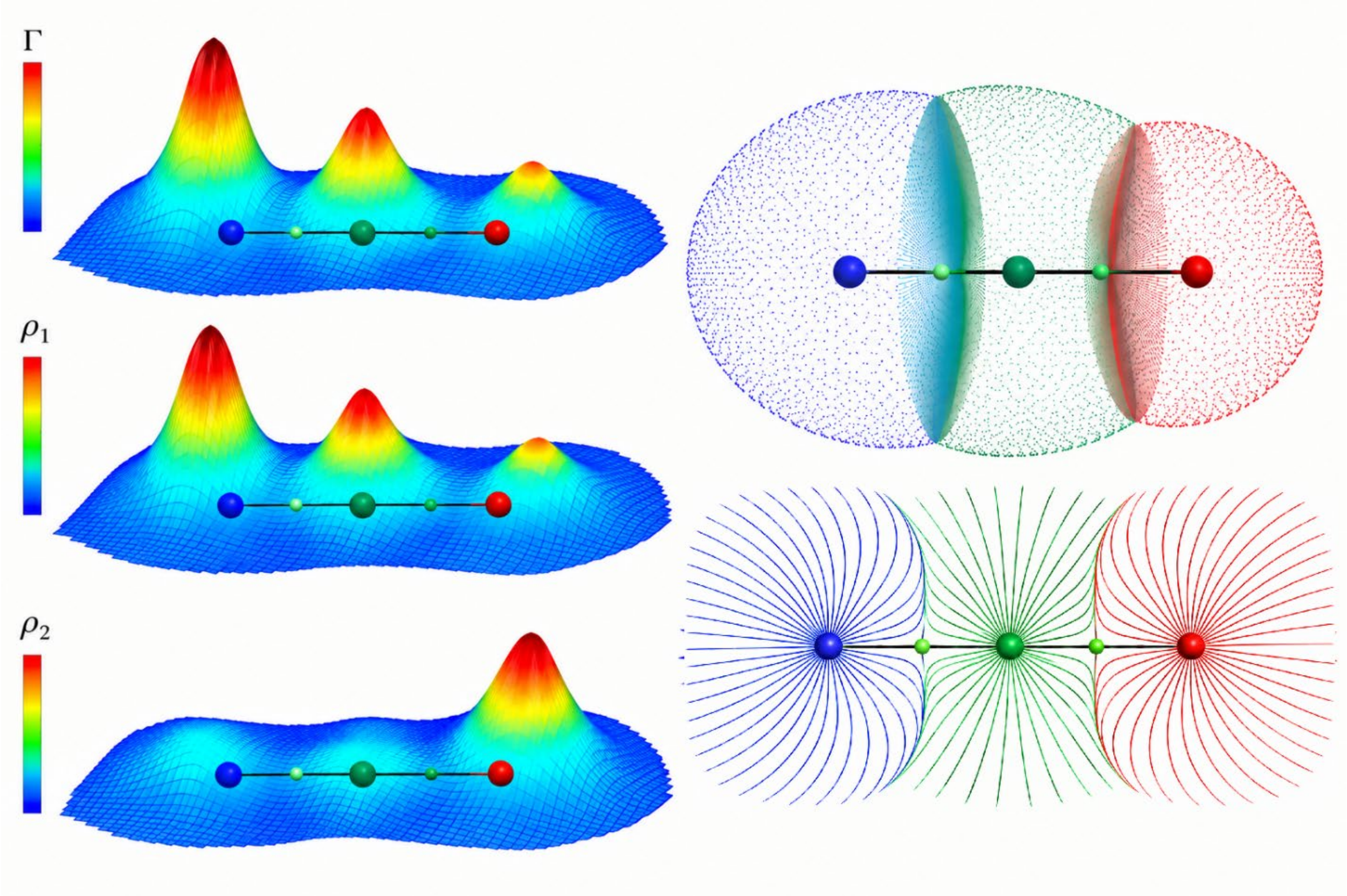

Figure 2- A schematic representation of the relief map of the Gamma density, the one-electron and one-nucleus densities, and the corresponding CPs for a three-atom molecule, shown on the left-hand side. In this depiction, electrons and the lightest nucleus are treated as quantum particles (see text for details). The outer surfaces of the atomic basins, represented as iso-surfaces of the Gamma density at $\Gamma^{(2)}(\vec{r}) = 0.001$ au, and the interatomic surfaces, defined as zero-flux surfaces of the Gamma density, are shown in the top right-hand side. The gradient vector field of the Gamma density in the plane containing all CPs is illustrated at the bottom right-hand side. The arrows of the gradient field have not been depicted for visual clarity, but they all point toward the CPs.

In addition to the three global attractors, two saddle CPs, classified as (3, -1) CPs, appear between each pair of global attractors. These CPs attract all remaining gradient paths that do not terminate at the global attractors and lie on the two zero-flux surfaces, which form the boundaries of the AIM. Two global attractors lie on or very near the clamped nuclei, while one lies in a basin that contains no clamped nucleus; if the nuclear density is localized, the basin encompasses most of that density. Nevertheless, this elegant concept—that the number of atomic basins equals the number of positively charged particles—does not apply to systems containing very light particles, such as positrons, which cannot accumulate enough electrons around themselves to form an independent basin. In a previous study, we demonstrated that for a positively charged particle with a single unit of positive charge, the particle's mass determines its ability to form an independent basin [108]. Below a critical mass, the exact value of which depends on the chemical environment, the positively charged particle cannot form a basin. Instead, the density of such particles resides within basins formed by heavier positively charged particles or clamped nuclei (see the next section for an example). All of these observations can be unified into a single mathematically rigorous conjecture, which is not addressed here (interested readers may consult Section 2.3 of [1] for details).

### 3.3. How to partition the total energy?

The fundamental property for any chemical partitioning analysis of molecular systems is the total energy [109]. In the context of MC-QTAIM, we examined various energy partitioning schemes, each with precedent in QTAIM [54,110,111]. Among these, the real-space dissection of the one- and two-particle terms of the MC Hamiltonian appears to be the "natural" approach [111]. This represents a generalized TC version of the interacting quantum atoms (IQA) scheme, originally developed by Pendas and coworkers for SC systems [112–114], and termed TC-IQA, which has been applied to positronic molecules [110,111]. In this subsection, I present the general formulation of MC-IQA as an illustrative application of the equations developed in the previous subsections. Nevertheless, before considering technical details, the aforementioned "naturalness" requires some rationalization. There are always "subjective" elements in any proposed energy partitioning method in quantum chemistry [62,115]. Hence, beyond the postulates of quantum mechanics, additional guiding principles are necessary to validate an energy partitioning scheme. The following principles have been taken into account in the proposed energy partitioning for MC systems:

1- Most chemical concepts, upon careful examination, can be extended beyond SC systems to MC systems, and an MC energy partitioning scheme serves this purpose. Therefore, any MC energy partitioning scheme must be translatable into chemical language, and the resulting energy terms should have clear chemical interpretations, as the ultimate goal is to quantify qualitative chemical concepts.
2- An MC energy partitioning scheme must provide meaningful explanations regarding the mechanisms underlying the formation of various bonds. While proposing a single, global mechanism for all bond types may be mathematically valid, it lacks chemical insight. Therefore, the scheme should distinguish between different bond types or interactions and offer specific, "localized" explanations for each.
3- It is preferable that the results of an MC energy partitioning scheme for a specific bond align with existing knowledge—namely, experimental, computational, and theoretical evidence—regarding that bond. If any contradictions arise, a clear explanation should be provided.
4- If applied to an SC system, the proposed MC energy partitioning scheme should reduce to one of the established SC energy partitioning schemes currently used for electronic systems; otherwise, it must produce chemically meaningful results for SC systems.

Since the MC Hamiltonian includes both single-particle and two-particle terms, it is necessary to use both one-particle and two-particle densities to partition the expectation value of the total energy. The partitioning of these terms at the initial stage is as follows:

$$E = \left\langle \Psi_{MC} \middle| \hat{H}_{MC} \middle| \Psi_{MC} \right\rangle = \sum_{p}^{S} \left[ \tilde{T}\left(\Omega_p\right) + \tilde{V}_{ext}\left(\Omega_p, \Omega_p\right) + \tilde{V}_{\mathrm{int}}\left(\Omega_p, \Omega_p\right) \right]$$

$$+ \sum_{p}^{S} \sum_{q \neq p}^{S} \left[ \tilde{V}_{ext}\left(\Omega_p, \Omega_q\right) + \tilde{V}_{\mathrm{int}}\left(\Omega_p, \Omega_q\right) \right] \quad (13)$$

Where the dissected intra- and inter-basin components—kinetic energy, external potential energy, $v_{ext}\left(\vec{r}_k\right) = \sum_{p}^{S} v_{ext}\left(\vec{r}_k, \Omega_p\right)$, and interaction energies—are expressed in the following forms:

$$\tilde{T}\left(\Omega_p\right)=\sum_k^M\left(-1/2m_k\right)\int_{\Omega_p} d\vec{r}_k \nabla_k^2 \rho_k^{(1)}\left(\vec{r}_k',\vec{r}_k\right)\Big|_{\vec{r}_k'=\vec{r}_k} ,$$

$$\tilde{V}_{ext}\left(\Omega_p,\Omega_p\right)=\sum_k^M\int_{\Omega_p} d\vec{r}_k \rho_k\left(\vec{r}_k\right) v_{ext}\left(\vec{r}_k,\Omega_p\right),\quad \tilde{V}_{ext}\left(\Omega_p,\Omega_q\right)=\sum_k^M\int_{\Omega_p} d\vec{r}_k \rho_k\left(\vec{r}_k\right) v_{ext}\left(\vec{r}_k,\Omega_q\right),$$

$$\tilde{V}_{\text{int}}\left(\Omega_p,\Omega_p\right)=\sum_k^M q_k^2\int_{\Omega_p} d\vec{r}_{k,1}\int_{\Omega_p} d\vec{r}_{k,2}\frac{\rho_{kk}^{(2)}\left(\vec{r}_{k,1},\vec{r}_{k,2}\right)}{\left|\vec{r}_{k,1}-\vec{r}_{k,2}\right|}+\sum_k^M\sum_{l>k}^M q_k q_l\int_{\Omega_p} d\vec{r}_{k,1}\int_{\Omega_p} d\vec{r}_{l,1}\frac{\rho_{kl}^{(2)}\left(\vec{r}_{k,1},\vec{r}_{l,1}\right)}{\left|\vec{r}_{k,1}-\vec{r}_{l,1}\right|}$$

$$\tilde{V}_{\text{int}}\left(\Omega_p,\Omega_q\right)=\sum_k^M q_k^2\int_{\Omega_p} d\vec{r}_{k,1}\int_{\Omega_q} d\vec{r}_{k,2}\frac{\rho_{kk}^{(2)}\left(\vec{r}_{k,1},\vec{r}_{k,2}\right)}{\left|\vec{r}_{k,1}-\vec{r}_{k,2}\right|}+\sum_k^M\sum_{l>k}^M q_k q_l\int_{\Omega_p} d\vec{r}_{k,1}\int_{\Omega_q} d\vec{r}_{l,1}\frac{\rho_{kl}^{(2)}\left(\vec{r}_{k,1},\vec{r}_{l,1}\right)}{\left|\vec{r}_{k,1}-\vec{r}_{l,1}\right|} \qquad (14)$$

The proposed partitioning is mathematically sound; however, it does not yet comply with the aforementioned principles and requires further refinement. In the next step, the typical decompositions of the reduced second-order density matrices and the joint densities into uncorrelated and correlated contributions are adopted, assuming that all components are fermions [111]:

$$\rho_{kk}^{(2)}\left(\vec{r}_{k,1},\vec{r}_{k,2}\right)=(1/2)\left[\rho_k\left(\vec{r}_{k,1}\right)\rho_k\left(\vec{r}_{k,2}\right)+\rho_{xc,k}\left(\vec{r}_{k,1},\vec{r}_{k,2}\right)\right],$$

$$\rho_{kl}^{(2)}\left(\vec{r}_{k,1},\vec{r}_{l,1}\right)=\rho_k\left(\vec{r}_{k,1}\right)\rho_l\left(\vec{r}_{l,1}\right)+\rho_{c,kl}\left(\vec{r}_{k,1},\vec{r}_{l,1}\right) \qquad (15)$$

In these expressions, "*x*" and "*c*" represent exchange and correlation, respectively. If a pair of particles, denoted by position vectors $\vec{r}_{k,1},\vec{r}_{k,2}$ or $\vec{r}_{k,1},\vec{r}_{l,1}$, repel each other, a Coulomb "hole" forms around each particle [116], and the correlated part of the density matrix describes this hole. Conversely, if the particles attract each other, the correlated part describes a Coulomb "hill," as recently discussed [117]. Incorporating all these effects into $\tilde{V}_{\text{int}}\left(\Omega_p,\Omega_q\right)$ in Eq. (13) yields:

$$E=\sum_p^S\left[\tilde{T}\left(\Omega_p\right)+\tilde{V}_{ext}\left(\Omega_p,\Omega_p\right)+\tilde{V}_{\text{int}}\left(\Omega_p,\Omega_p\right)\right]$$

$$+\sum_p^S\sum_{q\neq p}^S\left[\tilde{V}_{ext}\left(\Omega_p,\Omega_q\right)+\tilde{V}_{\text{int}}^{cl}\left(\Omega_p,\Omega_q\right)+\tilde{V}_{\text{int}}^{xc}\left(\Omega_p,\Omega_q\right)\right] \qquad (16)$$

Wherein:

$$\tilde{V}_{\text{int}}^{cl}\left(\Omega_p,\Omega_q\right)=(1/2)\sum_k^M q_k^2\int_{\Omega_p} d\vec{r}_{k,1}\int_{\Omega_q} d\vec{r}_{k,2}\frac{\rho_k\left(\vec{r}_{k,1}\right)\rho_k\left(\vec{r}_{k,2}\right)}{\left|\vec{r}_{k,1}-\vec{r}_{k,2}\right|}+\sum_k^M\sum_{l>k}^M q_k q_l\int_{\Omega_p} d\vec{r}_{k,1}\int_{\Omega_q} d\vec{r}_{l,1}\frac{\rho_k\left(\vec{r}_{k,1}\right)\rho_l\left(\vec{r}_{l,1}\right)}{\left|\vec{r}_{k,1}-\vec{r}_{l,1}\right|}$$

$$\tilde{V}_{\text{int}}^{xc}\left(\Omega_p,\Omega_q\right)=\left(1/2\right)\sum_k^M q_k^2\int_{\Omega_p} d\vec{r}_{k,1}\int_{\Omega_q} d\vec{r}_{k,2}\,\frac{\rho_{xc,k}\left(\vec{r}_{k,1},\vec{r}_{k,2}\right)}{\left|\vec{r}_{k,1}-\vec{r}_{k,2}\right|}+\sum_k^M\sum_{l>k}^M q_k q_l\int_{\Omega_p} d\vec{r}_{k,1}\int_{\Omega_q} d\vec{r}_{l,1}\,\frac{\rho_{c,kl}\left(\vec{r}_{k,1},\vec{r}_{l,1}\right)}{\left|\vec{r}_{k,1}-\vec{r}_{l,1}\right|} \quad (17)$$

The sum of the first three terms in Eq. (16) for each basin represents the intra-basin energy, $\tilde{E}_{\text{intra}}\left(\Omega_p\right)$, which is hereafter treated as a single entity, as in the IQA and TC-IQA schemes [110–114]. The difference between the intra-basin energy and the free atom energy, used as the reference energy, is commonly referred to as the basin "deformation" energy in the IQA/QTAIM literature [112–114]. However, AIM electron populations, $N_{ele}\left(\Omega_p\right)=\int_{\Omega_p} d\vec{r}\rho_{ele}\left(\vec{r}\right)$, usually differ from the number of electrons in the corresponding free atoms, and this imbalance introduces a "charge-transfer" (CT) energy contamination into the basin deformation energies [118–120]. In addition to CT, the "reorganization" of the internal structure of the AIM, induced by the chemical environment, also contributes to the deformation energy. Although disentangling these two contributions is not straightforward [118–120], it is preferable to refer to this combined effect as the CT-reorganization energy rather than simply deformation energy: $E_{\text{CT-reorg}}\left(\Omega_p\right)=E_{\text{intra}}\left(\Omega_p\right)-E_{ref}$. The next three terms in Eq. (16) correspond to various types of inter-basin interactions and can be categorized into two main contributions, electrostatic and exchange-correlation:

$$E=\sum_p^S\tilde{E}_{\text{intra}}\left(\Omega_p\right)+\sum_p^S\sum_{q\neq p}^S\tilde{E}_{\text{inter}}\left(\Omega_p,\Omega_q\right)$$

$$\tilde{E}_{\text{inter}}\left(\Omega_p,\Omega_q\right)=\sum_p^S\sum_{q\neq p}^S\left[\tilde{V}_{\text{elec}}\left(\Omega_p,\Omega_q\right)+\tilde{V}_{\text{int}}^{xc}\left(\Omega_p,\Omega_q\right)\right]$$

$$\tilde{V}_{\text{elec}}\left(\Omega_p,\Omega_q\right)=\tilde{V}_{ext}\left(\Omega_p,\Omega_q\right)+\tilde{V}_{\text{int}}^{cl}\left(\Omega_p,\Omega_q\right) \quad (18)$$

This is the final MC-IQA partitioning scheme, consistent with the four principles outlined above. The electrostatic contribution, $\tilde{V}_{\text{elec}}\left(\Omega_p,\Omega_q\right)$, includes all terms describing classical interactions between charge distributions of particles from the same or different components. If the external potential originates from point charges, each located within a basin, an additional term is added to the electrostatic contribution to account for inter-basin interactions between these static point charges. The electrostatic contribution in IQA is generally straightforward to interpret and encompasses interactions arising from various electrostatic multipole moments of AIM [121,122].

These multipole moments appear in the multipole expansion of each integral in the electrostatic contribution (see Chapter 5 in [10]). The simplest of these interactions is the charge-charge interaction, which underpins pure ionic bonding. If an MC system consists solely of electrons and heavy, localized positively charged particles, each confined to a single atomic basin, the nature of the electrostatic contribution from the MC-IQA analysis resembles that of the IQA. However, if the particles carry a negative charge or are light, delocalized positively charged particles, novel electrostatic interactions may arise; these represent new "resources" of nature for bonding. In a recent study, we demonstrated how these new resources bind two hydride anions, which would otherwise repel each other, through electrostatic interactions induced by a single positron [110] (see the next section for more details). In this species, the delocalized one-positron density is evenly distributed between the two atomic basins, and the one-positron density on one atom interacts electrostatically with the one-electron density of the other atom. This interaction produces a large stabilizing energy term that overcomes the electrostatic repulsion between the two negatively charged hydrides [110]. In addition to inter-basin electrostatic interactions, as evident from Eq. (18), the inter-basin exchange-correlation energy also contributes to the basin interaction energy. The exchange-correlation contribution in IQA is interpreted as the energetic manifestation of covalent bonding [123–125], although the pure correlation contribution is the source of van der Waals bonding [122]. In MC systems, each component contributes separately, as an "intra-component" term, to the exchange-correlation energy; however, for heavy, localized quantum particles, this contribution is less significant, and sometimes negligible, compared to that of light, delocalized particles. In positronic systems with two positrons exchanging between two atomic basins, this contribution highlights the positronic role in covalent bonding [111]. The most intriguing contribution arises from correlations between particles of two different components, termed the "inter-component" term; this represents a totally novel "resource" for bonding, absent in SC systems. In a recent study, we demonstrated that this term can contribute significantly to bonding in a species with two positrons [126] (see the next section for more details). The role of these novel contributions in bonding within other MC systems requires further investigation, and appropriate terminology must be developed for their classification. Additionally, as demonstrated in the TC-IQA analysis of positronic species, further subdivision of the electrostatic and exchange-correlation contributions into chemically informative subcomponents is feasible [111]. However, universal adoption across all MC systems will require additional case studies.

### 3.4. Q&A

1) *Is the MC-QTAIM formalism applicable to fully nonadiabatic ab initio nBO wavefunctions?*

Not in its current formulation. To apply MC-QTAIM, as with QTAIM, a "fixed frame" must be established to circumvent the "collective" translational and rotational motions and focus solely on the "internal" motions, such as electronic and vibrational motions. In QTAIM, this requirement is automatically satisfied because, in the SC Hamiltonian, the fixed charges naturally define a fixed frame within which only the internal motions of the lighter particles are described. The resulting one-particle and property densities derived from SC wavefunctions and used in the QTAIM analysis also possess this characteristic and are free from the effects of collective motions. More precisely, the collective motions are not eliminated but are instead contained within a separate Hamiltonian for the heavy components, such as the nuclear vibrational-rotational Hamiltonian. However, as emphasized in the response to the second question in the previous Q&A subsection, the wavefunction of the heavy particles does not influence the determination of atomic boundaries. Therefore, any basin contribution computed for the heavy components is added a posteriori to the atomic basins already established.

In contrast, for an MC Hamiltonian, the fixed frame is not guaranteed to persist. A fixed frame can be "designed" by choosing an appropriate form for $v_{ext}\left(\vec{r}_{i,k}\right)$ in the Hamiltonian in Eq. (1). This can be achieved by adding, for example, a set of "fictitious" harmonic oscillators for all or some of the particles, in addition to their physical mutual interactions. These harmonic traps localize some of the heavy particles around the fixed equilibrium points of the oscillators, thereby establishing the required fixed frame. This approach has been successfully applied to small few-body systems to eliminate translational motion [117,127–129]. In principle, it can also be extended to large MC systems, although perturbations induced by these fictitious traps on the physical motions of the particles are unavoidable. The eigenfunctions of this Hamiltonian can then be used for MC-QTAIM analysis.

If there are no external fields, $v_{ext}\left(\vec{r}_{i,k}\right)=0$ , or if the field is nonzero but does not impose a fixed frame, the MC Hamiltonian contains a mixture of collective and internal motions. This is the typical situation for the Hamiltonians used in ab initio nBO calculations. The resulting eigenfunctions and the derived one-particle and property densities also exhibit this characteristic, and they must be symmetric with respect to translations and rotations of the coordinate system. When these symmetries are imposed, no AIM structure emerges from the zero-flux equation of the Gamma or any other one-particle density. To address this issue, one must first separate the center-of-mass motion from the MC Hamiltonian by eliminating three degrees of freedom corresponding to the coordinates of the collective translational motion in space [42,130]. This yields a translation-free Hamiltonian with $3\left(\sum_{k=1}^{M} N_k - 1\right)$ variables that describe $\sum_{k=1}^{M} N_k - 1$ "pseudo-particles". Most ab initio nBO methods employ this Hamiltonian and solve the corresponding eigenvalue problem [40–42]. As mentioned, due to the residual collective rotational symmetry, no typical AIM structure is expected to emerge from the ab initio–derived translation-free eigenfunctions. Nevertheless, it is possible to eliminate an additional three degrees of freedom—Euler angles or similar rotational parameters—from the translation-free Hamiltonian to exclude collective rotations [130]. Unlike translational motion, rotational motion cannot be completely decoupled from internal motions; a coupling term invariably arises regardless of the chosen set of rotational variables [130]. The resulting internal (translation- and rotation-free) Hamiltonian is expressed in terms of $3\left(\sum_{k=1}^{M} N_k - 2\right)$ variables describing $\sum_{k=1}^{M} N_k - 2$ pseudo-particles. The axes of the coordinate system defined by these two transformations, commonly referred to as the body-fixed coordinate system, provide the fixed frame necessary for formulating the internal motion Hamiltonian, from which the eigenfunctions describing internal motions are derived [42]. It appears feasible that an extended version of the current MC-QTAIM, formulated to operate with these pseudo-particle variables, could perform AIM analysis using the internal eigenfunctions as input for simple few-body systems, such as the hydrogen molecule.

However, as Woolley emphasized in a detailed theoretical analysis that remains unchallenged [42,43,131–133], for more complex molecular systems with two or more isomers—such as hydrogen cyanide and hydrogen isocyanide—each nBO eigenfunction represents an exotic "superposition" of all isomers [48]. It is difficult to understand how a typical AIM structure can be attributed to such exotic wavefunctions that simultaneously describe multiple isomers. This issue, known as the "molecular structure conundrum," warrants thorough investigation in future studies [45,46].

2) *Does the issue of property density non-uniqueness in the QTAIM formulation also persist in the case of MC-QTAIM?*

Yes. The challenge arises from the fact that any variation of a property density is valid as long as it does not alter the expectation value of that property for the entire molecule. Let us assume we have a property density, $A_1(\vec{r})$, which, when integrated over all space, yields the expectation value: $\int_{R^3} d\vec{r} A_1(\vec{r}) = \langle \hat{A}_1 \rangle$. If one adds any function $f(\vec{r})$, satisfying the condition: $\int_{R^3} d\vec{r} f(\vec{r}) = 0$, to the property density, the new density: $A_1'(\vec{r}) = A_1(\vec{r}) + f(\vec{r})$, also yields the same expectation value. This implies that the sum of contributions over the entire basin is the same for both densities, $\sum_p^S A_1(\Omega_p) = \sum_p^S A_1'(\Omega_p)$, but the contribution of an individual atomic basin need not be equal: $\int_\Omega d\vec{r} A_1(\vec{r}) \neq \int_\Omega d\vec{r} A_1'(\vec{r})$. In a landmark paper [134], Ayers and coworkers derived an infinite number of kinetic energy density families, demonstrating this property explicitly for kinetic energy. In the original formulation of QTAIM [6], Bader implicitly restricted the kinetic energy densities to the Laplacian family: $f(\vec{r}) = \lambda \nabla^2 \rho_1(\vec{r})$ ($\lambda$ is a constant), which has the additional property: $\int_\Omega d\vec{r} f(\vec{r}) = 0$. However, as Ayers and coworkers rightly emphasized, this is an arbitrary choice.

I recently proposed a reformulation of QTAIM that circumvents this problem by working with "property distributions", $A_m(\Omega_k) = \binom{N}{m} \int_{\Omega_k} d\vec{r}_1 \ldots \int_{\Omega_k} d\vec{r}_m \int_{R^3-\Omega_k} d\vec{r}_{m+1} \ldots \int_{R^3-\Omega_k} d\vec{r}_N \sum_{spins} \Psi^* \hat{a}_1 \Psi$, instead of property densities [1]. In this new formulation, property distributions serve as the fundamental units for computing basin properties, $A(\Omega_P) = \sum_{m=1}^N m A_m(\Omega_P)$. Additionally, the variance of a property distribution quantifies fluctuations in basin properties, thereby conceptualizing basins as open subsystems [1]. Bader envisioned basin property fluctuations at a conceptual level [1,6]; however, in the original formulation of QTAIM, only electron number fluctuations between atomic basins were quantified [135–137]. This property distribution-based formalism has also been extended to MC-QTAIM (see Section 3.2 of [1] for details). Employing property distributions instead of property densities preserves the desired uniqueness of basin contributions.

## 4. When MC-QTAIM?

### 4.1. The broad picture

Based on the discussions in the previous sections, the answer to the question of when to apply MC-QTAIM may seem straightforward: as long as we have access to eigenfunctions of an MC Hamiltonian expressed in a fixed frame, regardless of the specific details of the system under study, these eigenfunctions can be used as inputs for MC-QTAIM analysis. This analysis identifies atomic basins—tightly packed clusters of particles in real space—and reveals the intra- and inter-cluster contributions to the total properties, as illustrated in Figure 1. Nevertheless, in practice, complexities arise due to the varying quality of currently available MC wavefunctions and limited access to software capable of performing MC-QTAIM analysis. I address some of these issues in the following subsections and briefly review the diversity of systems and quantum states amenable to MC-QTAIM analysis. Current applications of MC-QTAIM can be categorized into three main classes: (A) exotic molecular species, (B) exotic quantum states of conventional molecules, and (C) normal non-adiabatic states of conventional molecules. Classes A and B are briefly considered in this subsection, while class C is the focus of the next subsection.

Class A includes all systems in which certain parameters of the MC Hamiltonian—specifically, masses and charges—differ from those of electrons and the conventional nuclei composed of protons and neutrons. This class itself encompasses various subclasses (see Section 2.2 of [1] for details). The most intriguing species within this class, analyzed using MC-QTAIM, are the positronic species featuring one- and two-positron bonds [110,111,126,138–140]. In 2018, Reyes and coworkers made a landmark discovery through an ab initio computational study, demonstrating that a positron can bind two hydride ions, $\left[H^{-},e^{+},H^{-}\right]$, which otherwise repel each other [138]. This finding introduces a new dimension to chemical bond theory, as the chemical community traditionally considers electrons as the sole agents of bonding. Using high-level MC quantum Monte Carlo methods, Bressanini confirmed Reyes and coworkers' results [139]. In a subsequent study, he also showed that adding another positron to this system produces a new stable species, termed $\left(PsH\right)_2$ [140] (where $PsH$ denotes positronium hydride, an atom composed of two electrons, one positron, and a proton; for details, see [110,111]). We investigated the bonding modes in these two species using MC-QTAIM analysis [110,111]; Figure 3 schematically summarizes some of the results. Although both species exhibit typical homopolar diatomic structures, their bonding modes differ significantly. As discussed in Subsection 3.3, TC-IQA analysis reveals that the driving force behind the formation of the one-positron bond in $\left[H^-, e^+, H^-\right]$ is primarily electrostatic interaction arising from the interaction of the positron charge density in one atomic basin with the electron density in the other basin [110]. The origin of the two-positron bond in $(PsH)_2$ is more complex and remains under investigation [111,126,141]. While TC-IQA confirms that electrostatic interaction still plays a role in this species, it is insufficient to bind two charge-neutral $PsH$ atoms; instead, the binding results from stabilization induced by electron-positron and positron-positron correlations [126].

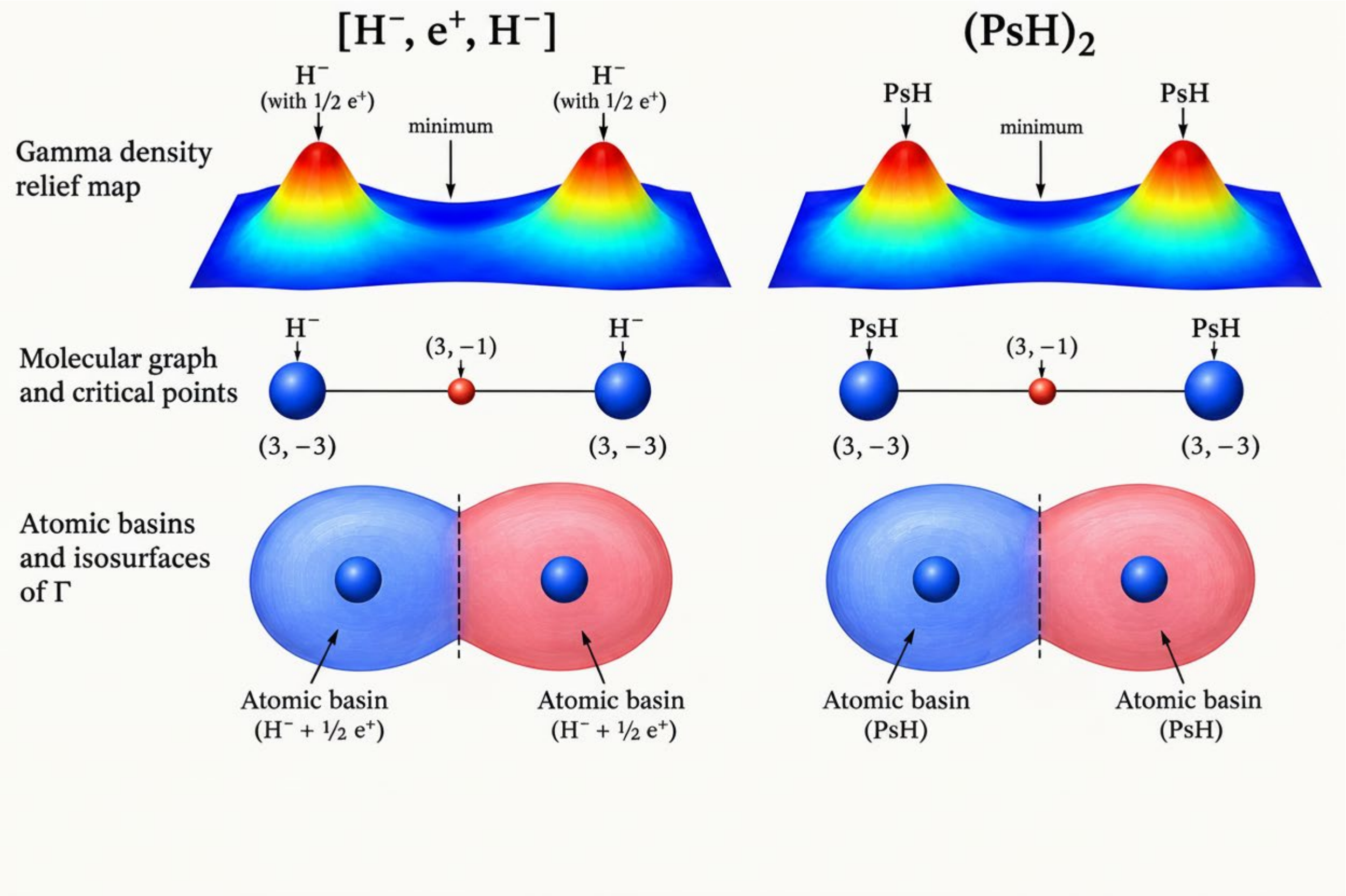

Figure 3- A schematic representation of the relief map of the Gamma density and the corresponding CPs, along with the outer surfaces of the two atomic basins in $\left[H^-, e^+, H^-\right]$ and $(PsH)_2$. In both species, each atomic basin contains half of the positron population, $N_{pos}(\Omega_p) = \int_{\Omega_p} d\vec{r} \rho_{pos}(\vec{r})$, and the molecular graphs of both systems resemble those of typical homopolar diatomic species (for computational details, see [110,111]).

The current ambiguity regarding the true nature of the two-positron bond stems from the lack of TC-IQA analyses employing high-quality, fully correlated TC wavefunctions. As discussed subsequently, the primary practical obstacle to applying MC-QTAIM and MC-IQA analyses is the limited availability of high-quality MC wavefunctions.

Class B includes all stationary quantum states characterized by an unusual delocalization of nuclei—particularly protons, the lightest nuclei—that exceeds what is predicted by the harmonic or anharmonic oscillator models of vibrations. Such states are commonly referred to as "superposed," and the corresponding one-nucleus density exhibits more than a single maximum in three-dimensional space, highlighting the exotic nature of delocalization [142–144].

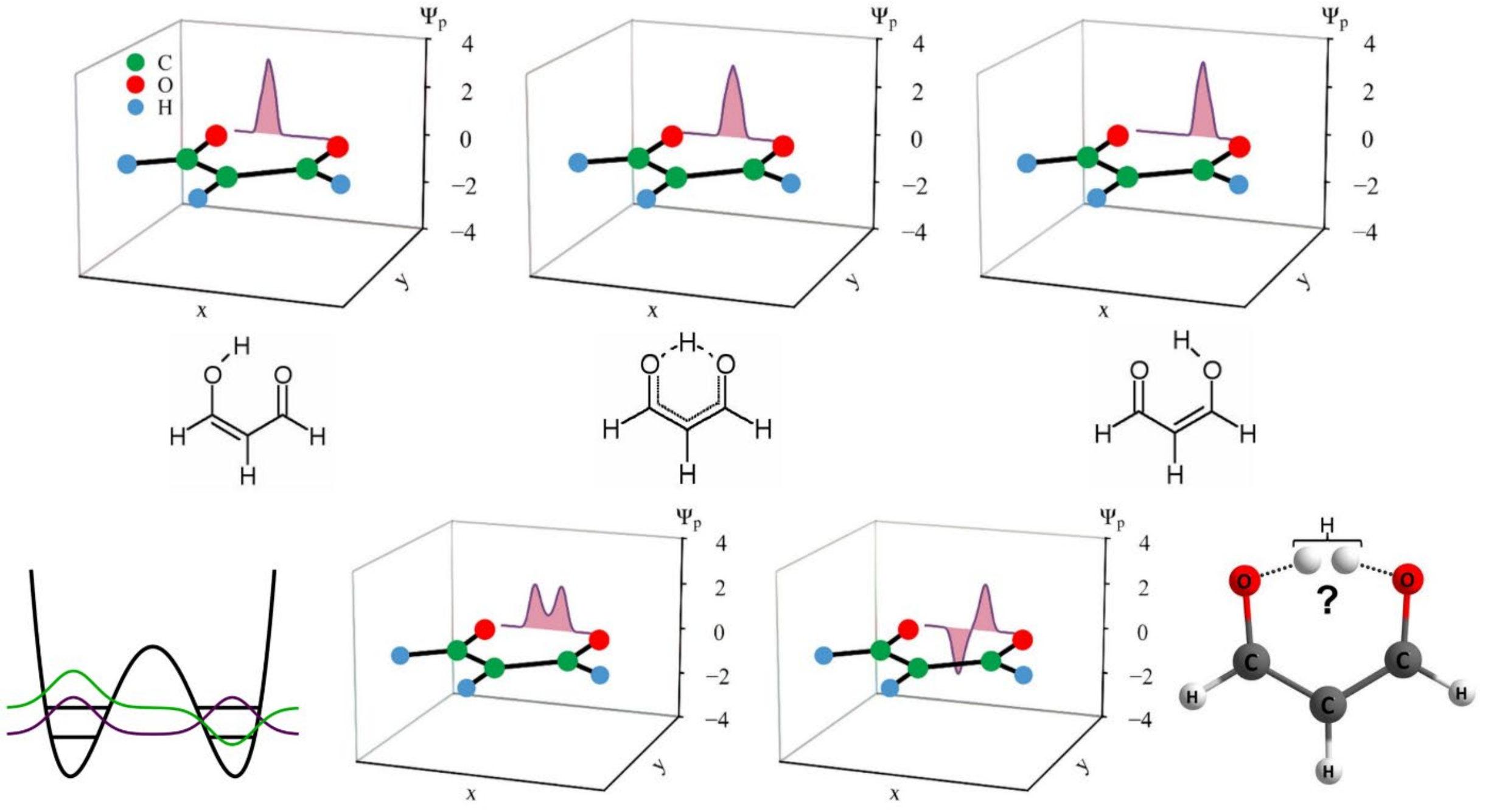

Figure 4- A schematic representation of the one-proton wavefunctions in the two conformers of malonaldehyde and the transition state between them, shown in the top row. These wavefunctions are part of an ab initio TC wavefunction, computed through partial optimization while artificially constraining the proton's motion around a specific point in space (for computational details, see [155]). In the bottom left, the ground and excited eigenfunctions of a particle in a double-well potential model are illustrated. These can be compared with the one-proton ground and excited wavefunctions shown in the middle of the row, which are also part of the ab initio TC wavefunction (for computational details, see [155]). The complex chemical structure of these delocalized states is schematically depicted in the bottom right.

The existence of such states has been theoretically demonstrated [43,44]; however, their experimental realization, although possible [145,146], is typically challenging due to environment-induced decoherence [147,148]. A straightforward example of superposed states arises in the model of a particle in a one-dimensional double-well potential, where the ground and certain excited states display the aforementioned double-maximum probability density (see Figure 4). If a nucleus in a real molecule experiences a double- or multiple-well potential energy surface, then the superposed state represents the true stationary state, as demonstrated in the case of the malonaldehyde molecule [149–154]. Figure 4 illustrates two localized, non-stationary, yet intuitively familiar one-proton wavefunctions for the proton situated between oxygen atoms, each corresponding to one of the two stable conformers of malonaldehyde (for technical details, see [155]). These localized one-proton wavefunctions can be understood as the usual vibrational states of the proton, in which the proton oscillates around a specific equilibrium position in space. The same figure also presents two delocalized, stationary, but exotic ground and excited one-proton wavefunctions, analogous to the double-well states depicted (for technical details, see [155]). Chemists typically conceptualize malonaldehyde as a dynamic chemical system in which the two conformers continuously interconvert via a transition state, shown in Figure 4 [156,157]. The primary driving force behind this dynamic behavior is the proton's quantum tunneling through the barrier separating the two conformers [158–163]. In a time-dependent process, the one-proton wavefunction, resembling a wave packet [164,165], oscillates between the two oxygen atoms. Inevitably, the hydrogen atom's basin also varies and oscillates between the oxygen basins; therefore, a time-dependent MC-QTAIM formalism is required to elucidate the details of this process, which currently does not exist [155]. The stationary states are more intriguing in this system, as the conventional concept of AIM is difficult to apply intuitively when the proton is not localized around a specific or equilibrium point in space (see Figure 4). The MC-QTAIM analysis of these superposed states, using ab initio TC wavefunctions in which electrons and the proton are treated as quantum particles within the external field generated by fixed nuclei, revealed that the hydrogen basin completely disappears [155]. Interestingly, replacing the proton with a positively charged muon restores an AIM structure similar to the transition state in malonaldehyde [166]. The chemical significance of these findings is still difficult to assess, and to the best of my knowledge, no one has explored molecular superposed states for chemical applications.

### 4.2. What does the MC-QTAIM analysis yield for usual molecules?

Before considering class C, it is timely to briefly review the ab initio orbital-based methodologies currently employed in MC quantum chemistry as the primary source of ab initio wavefunctions for MC-QTAIM. As discussed in Sections 1 and 2, when the MC Hamiltonian is used to describe a molecule—including all electrons and nuclei—the resulting nBO eigenfunctions are MC in nature. However, these wavefunctions are not amenable to MC-QTAIM analysis because of the obstacles detailed in the response to the first Q&A question in subsection 3.4. The simplest way to circumvent these obstacles is to treat only electrons and a limited number of light nuclei as quantum particles, while treating the heavy nuclei as fixed point charges (however, for alternative proposals, see [167–169]). The resulting Hamiltonian lies between the conventional SC electronic Hamiltonian, where all nuclei are fixed point charges, and the nBO Hamiltonian, which contains no fixed particles. Several research groups have attempted to extend the hierarchy of ab initio orbital-based electronic structure methods—originally developed to solve the electronic SE [8,9]—to address this type of MC Hamiltonian [170–175]. For a comprehensive review of the literature from the early 1970s through the first decade of the current century, including the classification of proposed wavefunctions, see Section 2 in [53]. The central idea behind this extension is to assign spin-spatial orbitals simultaneously to both electrons and light nuclei. The first rung of this hierarchy is the MC Hartree–Fock (MC-HF) method, which employs a wavefunction constructed as the product of Slater determinants for each fermionic component. In subsequent rungs, and consistent with electronic structure theory [8,9], correlations not captured by the MC-HF wavefunction are recovered using variational or perturbative methods, or through MC cluster expansions, by employing more sophisticated MC wavefunctions [170–175]. Within this hierarchy, in addition to familiar correlations between particles of the same component (e.g., electron–electron or proton–proton), new correlations arise between particles of different components (e.g., electron–proton) that are "qualitatively" distinct from the former [117]. This qualitative difference stems from the ubiquitous repulsion between charged particles within the same component, whereas attractive interactions can occur between charged particles from different components [117]. Efficiently recovering inter-component correlations remains a significant challenge for ab initio post-MC-HF methods [151]. This dilemma may only be resolved through novel theoretical developments beyond those already established in electronic structure theory [176–180]. An alternative to wavefunction-based MC ab initio methods is MC density functional theory (MC-DFT) [88,181], which explicitly incorporates correlation functionals for inter-component correlations [182–189]. Here, the challenge has shifted to efficiently designing inter-component correlation functionals [129]. These issues reveal that MC quantum chemistry is more a paradigm under development than a mature, well-established discipline like SC quantum chemistry. I will address some relevant issues that may affect the computational application of MC-QTAIM in the Q&A section.

The short answer to the title of this subsection is that MC-QTAIM analysis captures the "isotope effects" [95] when proper MC wavefunctions (see subsection 4.3 for details on what constitutes a proper wavefunction) are obtained for a molecular system and its isotopic substituents. As an illustrative example, replacing the mass of the proton(s) with that of the deuterium(s) in the MC Hamiltonian yields a new ab initio MC wavefunction for the isotopically substituted molecule. As mentioned in the second question in the Q&A of subsection 2.2, the nuclear mass may affect the basin properties derived from QTAIM analysis by adding a new contribution to the already computed electronic contribution. In contrast, within MC-QTAIM, since atomic boundaries are derived from the outset through the Gamma density, the basin boundaries also vary with changes in the MC wavefunction after isotopic substitution. This arises from the fact that not only do one-proton and one-deuterium densities differ because of the mass difference, but the nuclear mass also affects, albeit slightly, the one-electron densities, which differ between the protonic and deuteronic species. Additionally, the one- and two-electron property densities also differ slightly between the original system and its isotopic substituent. This means that the intra- and inter-basin properties for the isotopically substituted basin must differ from those of the original basin. These slight variations in electronic contributions represent the non-adiabatic effects discussed previously and are captured by MC wavefunctions. Broadly speaking, basin properties are a combination of adiabatic (already captured by QTAIM analysis) and non-adiabatic electronic and nuclear zero-point contributions; thus, isotopic effects are "naturally" included in MC-QTAIM analysis. As a proof of concept, we computed MC-HF wavefunctions to deduce TC wavefunctions for *HCN* and its deuterium-substituted derivative, *DCN*, and then performed MC-QTAIM analysis on both species [57]. In MC-HF calculations, the carbon and nitrogen nuclei were treated as fixed point charges, and their geometry, as well as the spin-orbitals for proton/deuterium and surrounding electrons, were fully optimized. Although this level of theory is not reliable for quantitative analysis (see subsection 4.3 for details on what quantitative means), it may semi-quantitatively convey the results expected from such analysis. The qualitative features of the one-electron, one-proton/deuterium, and Gamma densities are similar to those in Figure 2, and both species have three atomic basins. Indeed, the computed atomic charges, electron delocalization indices, and inter-basin electronic exchange energies differ slightly between the two species, revealing the expected non-adiabatic effects on the electronic contributions (for detailed definitions of these properties and numerical results, see [57]). Given the importance of deuterium substitution in modern medicinal chemistry and drug discovery [190,191], MC-QTAIM analysis could serve as a valuable new tool in the computational toolbox for studying relevant molecules.

### 4.3. Q&A

1) *To what extent are the ab initio wavefunctions computed using currently available orbital-based MC quantum chemical methods reliable for MC-QTAIM analysis of conventional molecules (Class C)?*

Not in general. Let us focus on certain typical problems associated with ab initio wavefunctions of TC systems composed of electrons and protons; generalization to MC systems, including heavier nuclei, is straightforward. To obtain an authentic MC wavefunction for MC-QTAIM, the primary issue—as discussed in the answer to question 1 of Q&A in subsection 3.4—is to fix the proper number of heavy nuclei to prevent translational and rotational contamination of the wavefunctions (some of our earlier computational studies violate this condition and should be regarded merely as "proof of concept" studies [53,55,102,192,193]). This condition implies that diatomic hydride systems are not amenable to MC-QTAIM analysis, and for larger systems, heavy-nucleus fixing must be performed carefully to avoid rotational contamination.

The primary challenge with orbital-based ab initio MC wavefunctions is the "over-localization" of the computed one-proton densities, which arises from the complete absence of electron-nucleus correlation within the MC-HF wavefunction. This issue is not easily resolved, even in post-MC-HF wavefunctions [194,195]. Such over-localization prevents accurate quantitative predictions of proton-related observables, such as vibrational frequencies and proton affinities [188,196]. Semi-empirically designed electron-proton correlation functionals, currently employed with MC Kohn-Sham (MC-KS) equations, may partially address this problem [182–189]. However, these functionals typically fail to predict all critical properties—such as one-proton densities and total energies—with consistent accuracy [187]. Unlike the Levy-Perdew first-principles framework for designing electronic exchange-correlation functionals [197,198], no systematic approach has yet been proposed for developing improved electron-proton correlation functionals. Nonetheless, in my view, MC-DFT remains the most promising avenue for practical MC quantum chemistry applicable to large molecules. A well-designed TC quantum chemical method must accurately describe the coupled electron-proton motion and recover the associated subtle nonadiabatic contributions to observables, which are generally much smaller than the dominant adiabatic effects [199]. We are currently far from achieving this ideal, and existing electron-proton correlation functionals are primarily tuned to reproduce adiabatic effects correctly. Moreover, in a TC framework describing electrons and protons, the vibrational motion of protons inevitably decouples from that of the other nuclei, which is generally an inadequate approximation. The proton-to-heavier-nuclei mass ratio, often used to justify the adiabatic approximation, is not as small as the electron-to-nuclei mass ratio. Consequently, there must be significant coupling between the motion described by the TC Hamiltonian and that governed by the Hamiltonian of the heavier nuclei. Although methods have been developed to compute this coupling, they introduce additional complexity to MC quantum chemistry [200–202]. An alternative approach, which we recently advocated, is to restrict the TC description to systems where the normal vibrational modes of the protons included in the TC treatment are sufficiently decoupled from the remaining normal modes [189].

How do these factors affect the MC-QTAIM analysis? When delineating AIM boundaries, the over-localization problem may have a relatively minor effect because the one-proton density contributes little to the Gamma density near the AIM boundaries (see Figure 2). However, over-localization can significantly impact basin properties. In TC-IQA analysis, over-localization artificially increases the kinetic energy of protons and the associated basin energy terms. Moreover, an over-localized one-proton density produces an artificially larger electrostatic interaction with the one-electron density, affecting both $\tilde{E}_{\text{intra}}\left(\Omega_p\right)$ and $\tilde{E}_{\text{inter}}\left(\Omega_p,\Omega_q\right)$. Although these pathological features may have a less pronounced effect on the pure electronic contributions to electrostatic and exchange-correlation energies, the TC-IQA analysis with over-localized wavefunctions is generally unreliable. The impact of incorporating vibrational coupling with heavier nuclei into MC-QTAIM analysis is more subtle, and no relevant theoretical or computational studies have been published to date.

Taking all of this into account, it seems reasonable to limit the MC-QTAIM analysis of conventional molecules to MC-KS wavefunctions derived using an appropriate electron-proton correlation functional. I recommend using MC-QTAIM analysis of MC-HF wavefunctions only as an intermediate reference state to elucidate the role of electron-proton correlation. Additionally, if vibrational couplings exist between the proton(s) and other nuclei, the analysis may exhibit unphysical features (one of our earlier computational studies may suffer from the same issue and should be regarded solely as a “proof of concept” study [203]). As computer scientists say, "garbage in, garbage out".

*2) Is there dedicated software available for automatically performing MC-QTAIM analysis?*

No, we have used our build-at-home version for computational studies; however, let me consider the details for possible future developments. To perform MC-QTAIM analysis, we need to prepare MC wavefunctions, which requires ab initio MC calculations on the target molecules. The resulting ab initio or MC-KS wavefunctions are then saved in one of the standard formats, such as .wfn or .wfx (see: https://aim.tkgristmill.com/wfxformat.html). These files are subsequently used as inputs for computational packages that perform MC-QTAIM analysis.

When we began developing MC-QTAIM, the primary obstacle to perform MC-QTAIM analysis was the lack of ab initio MC quantum chemical software packages, necessitating the development of custom programs from scratch. However, in recent years, many well-established quantum chemical computational packages have incorporated elements of the Nuclear-Electronic Orbital (NEO) MC hierarchy [204–209], originally developed by Hammes-Schiffer and colleagues [170,173–175], into their core functionality. Additionally, the LOWDIN package, developed by Flores-Moreno, Reyes, and collaborators, has been specifically designed for ab initio MC calculations [171,172]. Consequently, the current challenge lies in establishing a standardized protocol for saving MC wavefunctions in .wfn or .wfx file formats. Fortunately, with minor modifications to existing standard formats, it is possible to save an MC wavefunction. For MC-HF or MC-KS wavefunctions, the supporting information includes a sample .wfn file for the *HCN* species, as described previously [57]. In the extended format, the mass of components, their charge content, and the order of occupied orbitals are added as a separate header, as follows:

*HF-NEO [6-311+G(d,p): 4s1p/1s]*

*m = 1.0, m = 1836.15267*

*q = -1.0, q = +1.0*

*MOL 1-7, 8*

The first line provides information about the level of the ab initio calculation; the notation used for the basis set is detailed in the original paper [57]. The second line specifies the masses of the two components—the light component (component 1) and the heavy component (component 2)—in atomic units. The third line indicates the charge contents of these components in the same order as given in the second line. The fourth line identifies the orbital order, where the first seven orbitals belong to component 1 and the last orbital to component 2. The remainder of the file closely resembles a standard .wfn file, containing information about basis functions—some located on fixed nuclei and others on a Banquet (Bq) center that describes the protonic orbital and the electronic orbital of surrounding electrons. Adding more components does not introduce additional complexity; the new information can be incorporated into the second through fourth lines. Quantum chemical packages that perform NEO calculations can readily implement automatic generation of these proposed file formats, and users can initiate this process by including an extra command in the computation instructions.

If the appropriate file formats are available, it would be straightforward to modify existing packages for QTAIM analysis to perform MC-QTAIM analysis. The linear nature of density combination in MC-QTAIM ensures that the same modules used for topological analysis or the integration of property densities in QTAIM analysis [210–213] (see: https://aim.tkgristmill.com/references.html) can be adapted for MC-QTAIM with minimal effort, simply by duplicating the modules and designing a straightforward combinator. Implementing this would significantly expand the widespread applications of MC-QTAIM analysis.

## 5. Conclusions

The MC quantum chemistry is a rapidly growing field with numerous potential applications to both conventional and exotic molecules, as partially outlined in this review. However, its primary strength does not lie in competing with the SC quantum chemistry and the associated electronic structure theories in domains where these methods perform well. As demonstrated in a recent study using a toy model [117], when the non-adiabatic contribution to energy (or other observables) is smaller than the inter-component correlation energies, the conventional BO perturbative approach or BH-based variational methods are preferable, and MC ab initio methods offer limited advantages. This scenario typically applies to conventional molecules under ambient pressure conditions. The situation changes dramatically when the mass of the quantum particles added to the electronic system approaches that of the electron or when high hydrostatic pressures significantly couple electronic and vibrational states. This suggests that for positronic molecules, which have recently become common in biological and medical contexts [70,76,214], MC quantum chemical methods represent viable options for ab initio calculations. Additionally, high-pressure chemistry—with its many intriguing condensed phases and phenomena [215–217], such as metallic hydrogen [218–220], and high-temperature hydride superconductors [221,222]—provides a promising domain for the application of MC quantum chemistry. The complex nature of bonding in species and phases within these two areas [223,224] makes the application of MC-QTAIM particularly informative and valuable. Perhaps this will help, at least partially, to realize Hefferlin's dream of extending chemical concepts to areas and types of matter that are unfamiliar to conventional chemistry [63].

**Acknowledgements**

I would like to thank Dr. Payam Nasertayoob and Dr. Mohammad Goli for their significant contributions to the development of MC-QTAIM. My initial collaboration with Dr. Nasertayoob, which involved reconsidering the foundations of QTAIM, paved the way for its subsequent extension to MC systems. Dr. Goli deserves full credit for the primary theoretical advancements of MC-QTAIM and the computer programming necessary to apply the method to real molecules; without his efforts, none of the results discussed in this paper would have been possible. This work was supported by the HESC of Armenia under grant No. 25IRF-1C002.